 \documentclass[twocolumn]{aastex631}
\usepackage{amsmath}
\usepackage{breqn}
\usepackage{graphicx}

\font\manual=manfnt at 7pt \def\dbend{\hbox{\raise0.9ex\hbox{\manual\char127\hspace{0.6em}}}}

\providecommand{\e}[1]{\ensuremath{\times 10^{#1}}}

\newcounter{INTERNALionstage}

\def\gtsim{\mathrel{\hbox{\rlap{\hbox{\lower4pt\hbox{$\sim$}}}\hbox{$>$}}}}
\def\lesssim{\mathrel{\hbox{\rlap{\hbox{\lower4pt\hbox{$\sim$}}}\hbox{$<$}}}}

\def\A{{\rm\thinspace \AA}}

\def\g{{\rm\thinspace g}}

\def\K{{\rm\thinspace K}}

\def\m{{\rm\thinspace m}}

\def\s{{\rm\thinspace s}}

\def\W{{\rm\thinspace W}}

\def\oviii{\mbox{{\rm O~{\sc viii}}}}

\def\sixiii{\mbox{{\rm Si~{\sc xiii}}}}
\def\sixiv{\mbox{{\rm Si~{\sc xiv}}}}

\def\sxv{\mbox{{\rm S~{\sc xv}}}}

\def\arxvii{\mbox{{\rm Ar~{\sc xvii}}}}

\def\caxix{\mbox{{\rm Ca~{\sc xix}}}}

\def\nixxvii{\mbox{{\rm Ni~{\sc xxvii}}}}
\def\nixxvi{\mbox{{\rm Ni~{\sc xxvi}}}}
\def\h0{\mbox{{\rm H}$^0$}}
\DeclareMathAlphabet{\vib}{OML}{cmm}{m}{it}

\renewcommand{\thefootnote}{\fnsymbol{footnote}}

\shorttitle{Chemical enrichment in A2029}
\graphicspath{{./}{figures/}}

\begin{document}

\title{
Unveiling Chemical Enrichment in the Abell 2029 Core with XRISM, XMM-Newton, and Chandra}

\author[0000-0002-5222-1337]{Arnab Sarkar}
\affiliation{Kavli Institute for Astrophysics and Space Research,
Massachusetts Institute of Technology, 70 Vassar St, Cambridge, MA 02139, USA}
\affiliation{Department of Physics, University of Arkansas, 825 W Dickson st.,
Fayetteville, AR 72701, USA}
\email{arnabsar@mit.edu, arnabs@uark.edu}

\author[0000-0002-3031-2326]{Eric D.\ Miller}
\affiliation{Kavli Institute for Astrophysics and Space Research,
Massachusetts Institute of Technology, 70 Vassar St, Cambridge, MA 02139, USA}

\author{Brian McNamara}
\affiliation{Department of Physics \& Astronomy, Waterloo Centre for Astrophysics, University of Waterloo, Ontario N2L 3G1, Canada}

\author[0000-0001-5880-0703]{Ming Sun}
\affiliation{Department of Physics and Astronomy, The University of Alabama in Huntsville, Huntsville, AL 35899, USA}

\author[0000-0002-7962-5446]{Richard Mushotzky}
\affiliation{Department of Astronomy, University of Maryland, College Park, MD 20742, USA}

\author[0000-0003-4117-8617]{Stefano Ettori}
\affiliation{INAF, Osservatorio di Astrofisica e Scienza dello Spazio, via Piero Gobetti 93/3, 40129 Bologna, Italy}
\affiliation{INFN, Sezione di Bologna, viale Berti Pichat 6/2, 40127 Bologna, Italy}

\author[0000-0002-3754-2415]{Lorenzo Lovisari}
\affiliation{INAF, Istituto di Astrofisica Spaziale e Fisica Cosmica di Milano, via A. Corti 12, 20133 Milano, Italy}
\affiliation{Center for Astrophysics $|$ Harvard $\&$ Smithsonian, 60 Garden Street, Cambridge, MA 02138, USA}

\author[0000-0001-7630-8085]{Irina Zhuravleva}
\affiliation{Department of Astronomy and Astrophysics, University of Chicago, Chicago, IL 60637, USA}

\author[0000-0002-2784-3652]{Naomi Ota}
\affiliation{Department of Physics, Nara Women's University, Nara 630-8506, Japan}



\begin{abstract}
We present new measurements of
the chemical abundance pattern in the core of the nearby galaxy cluster 
Abell~2029, based on XRISM
observations with Resolve
(37 ks) and Xtend (500 ks), 
combined with archival data from 
XMM-Newton (EPIC, RGS) and Chandra.
Fe abundances derived from Resolve, 
Xtend, and EPIC are broadly consistent, 
while RGS gives systematically lower values.  
Because the XRISM gate valve remained closed during these observations, 
Resolve spectral fitting is restricted to
the 2--10 keV band,
providing reliable constraints only for
elements with strong lines in this band 
(S, Ar, Ca, Fe, Ni). 
Abundances of the $\alpha$-elements are therefore derived using
complementary observations 
from Xtend, EPIC, RGS, and Chandra.
We construct an average X/Fe pattern
in the cluster core by using Resolve 
exclusively
for S/Fe, Ar/Fe, Ca/Fe, and Ni/Fe, 
and RGS + Xtend for O/Fe.
The Ne/Fe ratio is averaged from Xtend, EPIC, RGS, and Chandra
measurements;
Mg/Fe from EPIC and Chandra measurements; 
and Si/Fe 
from Xtend, EPIC, and Chandra.
Comparison with the supernovae yield
models indicates that the observed 
abundance pattern in A2029 core is
best reproduced 
by a combination of core-collapsed
yields from
low-metallicity progenitors 
($Z_{\rm init}=0.001$) and a 
sub-Chandrasekhar-mass, double-degenerate 
Type Ia model. 
Additionally,
we find an excess in Ca abundance 
in the core of A2029 that cannot be
reproduced by the standard supernovae yield models. 

\end{abstract}


\section{Introduction}
High-resolution X-ray spectroscopy 
serves as a powerful tool to precisely measure the chemical abundances of the 
intra-cluster medium (ICM) in galaxy clusters. 
The observed emission lines primarily arise from K- and L-shell transitions
in highly ionized, hot plasma \citep[e.g.,][]{2020ApJ...901...68C,2020ApJ...901...69C,2017A&A...607A..98D,2019MNRAS.483.1701S,2023ApJ...953..112F}.
The Resolve X-ray microcalorimeter on 
board XRISM 
\citep{2020SPIE11444E..22T,2022SPIE12181E..1SI},
with its $\sim4.5$ ev-level 
spectral resolution, enables high-fidelity and precise measurements of ICM chemical abundances
by resolving individual K-shell
emission lines.
However, under the gate-valve-closed configuration, 
the practical energy band for extended 
sources is limited to $>2~\mathrm{keV}$.
Consequently, 
metal abundance measurements by Resolve
are currently restricted to the He- and H-like 
$K\alpha$ complexes of Si, S, Ar, Ca, Cr, Mn, Fe, and Ni (i.e., Fe-peak elements).
Within this band, Resolve’s 
eV-level resolving power cleanly
separates blends in the weak 
Fe-peak K lines (Cr, Mn, Ni) that 
are otherwise blended with Fe features, allowing robust Fe-peak/Fe abundance ratios
that directly probe Type Ia supernova
enrichment channels. 
Several XRISM Performance Verification (PV)-phase observations of 
nearby 
clusters confirm strong detections
of these K-shell lines 
\citep[e.g.,][]{A2029_eric,A2029_naomi,A2029_Sarkar,2025arXiv250805067X,2025PASJ..tmp...89F,2025ApJ...985L..20X}.

With a better understanding of stellar nucleosynthesis over the past
decades, it is now evident that most of the lighter elements, like O,
Mg, and Ne, are synthesized by massive stars
($\sim10-140\ M_\odot$) and expelled into the
ICM by core-collapse supernovae (SNcc; e.g., \citealt{2006NuPhA.777..424N,2013ARA&A..51..457N}).
On the other hand,
the heavier elements, like Ar, Ca, Fe, and Ni, are primarily
produced by Type Ia supernovae (SNe Ia; e.g., \citealt{1999ApJS..125..439I,2016A&A...595A.126M,2022MNRAS.516.3068S}).
For decades, the SNIa yield models have predominantly focused on scenarios
involving a carbon–oxygen white dwarf (WD)
that approaches the Chandrasekhar limit
either through stable hydrogen-rich
accretion from 
a non-degenerate companion (single degenerate; \citealt{1973ApJ...186.1007W}), or through the violent merger of two C/O WDs (double degenerate; \citealt{1984ApJS...54..335I}).
In such models, central carbon ignition 
initiates a simmering phase, which is 
subsequently followed by the onset of a deflagration, a transition to detonation, 
and ultimately a thermonuclear explosion 
\citep[e.g.,][]{2004ApJ...612L..37P,2013MNRAS.429.1156S}.

More developments from last decade have shown
that helium-accreting systems,
particularly those involving helium white
dwarf donors, can lead to significantly 
smaller helium shells at ignition due to
higher accretion rates \citep{2007ApJ...662L..95B,2009ApJ...699.1365S}. 
These insights have led to a renewed 
interest in sub-Chandrasekhar-mass (sub-M$_{\rm Ch}$) explosions, 
particularly in the context of 
double-detonation models 
\citep[e.g.,][]{2010A&A...514A..53F,2010ApJ...719.1067K,2011ApJ...734...38W,2014ApJ...785...61S}. 
Parallel studies of dynamically unstable
WD mergers have also suggested that
helium detonation may be triggered
during the merger phase itself, 
as mass is rapidly transferred 
\citep[e.g.,][]{2010ApJ...709L..64G,2013ApJ...770L...8P}. 
This pathway has gained further traction,
but it is not
yet fully understood.
While the nature of SN Ia progenitors
remains a long-standing open question in astrophysics, 
it is evident that the detailed abundance patterns of the metals
produced are intimately linked to 
the physics of the explosion mechanism \citep[e.g.,][]{2012PASA...29..447M,2014ARA&A..52..107M,2018ApJ...854...52S,2025A&ARv..33....1R}.

Galaxies are open systems,
gaining pristine gas and 
losing enriched gas, making their remaining metallicity highly model dependent.
Metals produced within cluster 
member galaxies accumulate in the
ICM over very long periods $(>5$ Gyr) and represent a true average over many individual galaxies.
The total mass of metals deposited in the cluster core is approximately 2–6 times greater than the metal mass retained within the galaxies themselves \citep{2014MNRAS.444.3581R,2018SSRv..214..129M,2021Univ....7..208G,2024A&A...685A..88M}. 
As a result, the metal abundances in the ICM serve as a “fossil” record of the 
cumulative yields from successive generations of stars, 
released through supernova explosions and stellar winds. 
These abundance patterns offer a unique
test bed to constrain
SNcc and SNIa yield models as well as other aspects of star formation, IMF and galactic winds. 

This paper 
(Paper IV of the Abell~2029 XRISM PV observation series), 
 focuses on the 
chemical enrichment in
the core of Abell~2029 galaxy cluster.
To constrain the abundances of 
low-energy $\alpha$-elements, 
we supplement the Resolve data with simultaneous XRISM Xtend observations and with 
archival observations from XMM-Newton RGS
and EPIC, as well as Chandra ACIS. 
Together, we derive elemental abundances for nine elements:
O, Ne, Mg, Si, S, Ar, Ca, Fe, and Ni.  
Throughout this work, we adopt a 
$\Lambda$CDM cosmology with
\( H_0 = 70 \ \, \text{km s}^{-1} \, \text{Mpc}^{-1} \),
\( \Omega_m = 0.3 \), 
and \( \Omega_\Lambda = 0.7 \).
At the redshift of Abell~2029 
(\( z = 0.0787 \)), 1$'$
corresponds to 89 kpc. 
We adopt the proto-solar abundance table of \citet{Lodders09} throughout. 
Unless otherwise noted, all uncertainties 
are reported at the
1\( \sigma \) (68\%) 
confidence level.

\section{Data reduction}

\begin{table*}
\caption{Observation logs\label{tab:obs_log}}   
\begin{center}
\setlength{\tabcolsep}{3pt}
\begin{tabular}{cccccccc}
Telescope & Instrument$^{\dagger}$ & Pointing & Observation ID & RA & Decl. & Observation Date & Exposure\\
& & & & (deg) & (deg) & & (ks)\\
\hline
\hline
& Resolve/Xtend & Central & 000149000 & 227.7341 & 5.7451 & 2024-01-10 & 12.44\\
& Resolve/Xtend & Central &  000151000 & 227.7331 & 5.7450 & 2024-01-13 & 25.11\\
XRISM & Xtend & North 1 & 000150000 & 227.7635 & 5.7850 & 2024-01-10 & 106.03\\
& Xtend & North 2 & 300053010 & 227.7943 & 5.8245 & 2024-07-27 & 366.15\\
\hline
& PN/MOS/RGS & $-$ & 0111270201 & 227.7333 & 5.7450 & 2002-08-25 & 23.5\\
& PN/MOS/RGS & $-$ & 0551780201 & 227.7447 & 5.7616 & 2008-07-17 & 46.8\\
XMM-Newton & PN/MOS/RGS & $-$ & 0551780301 & 227.7447 & 5.7619 & 2008-07-19 & 46.8\\
& PN/MOS/RGS & $-$ & 0551780401 & 227.7447 & 5.7616 & 2008-07-21 & 46.8\\
& PN/MOS/RGS & $-$ & 0551780501 &  227.7447 & 5.7616 & 2008-07-18 & 33.5\\
\hline
& ACIS-S & $-$ & 891 & 227.7267 & 5.7636 & 2000-04-12 & 19.8\\
Chandra & ACIS-I & $-$ & 6101 & 227.7632 & 5.7636 & 2004-12-17 & 9.9\\
& ACIS-S & $-$ & 4977 & 227.7469 & 5.7582 & 2004-12-17 & 76.8\\
\hline
\end{tabular}
\end{center}
$^{\dagger}$ Instrument data analyzed in this
paper.
\end{table*}

\subsection{XRISM/Resolve}\label{sec:rslv_data_reduction}
XRISM observed the central 
region of Abell 2029 through
two co-aligned pointings on 
2024 January 10 and 13, with
exposure times of 12.44 ks 
(OBSID 000149000) and 
25.11 ks (OBSID 000151000), respectively. 
Two additional offset pointings
were conducted: North 1, observed between 2024 January 
10–13 for 106.03 ks 
(OBSID 000150000), and North 2, observed between 2024 July 27 
and August 4 for 366.15 ks 
(OBSID 300053010).
A complete observation log is
provided in Table \ref{tab:obs_log}.
In this work, we utilize data 
from both the Resolve and Xtend instruments.
The data were reprocessed 
using HEASoft v6.34 and the most recent calibration database 
(CalDB v20250915), following the standard screening procedures
adopted in the XRISM analysis
of N132D \citep{XRISM2024_N132D}.

For detailed data reduction 
methods for Resolve,
we refer readers to  Paper I 
\citep{A2029_eric}, Paper II 
\citep{A2029_naomi}, and
Paper III \citep{A2029_Sarkar}.
In summary, energy-scale drift in Resolve during observations is
tracked by periodically illuminating the focal plane with a built-in $^{55}$Fe 
source mounted on the filter wheel, typically during Earth occultation. 
After gain reconstruction,
the flat-field-averaged energy scale uncertainty is estimated to be 
$\pm 0.3$ eV in the 5.4–9 keV band, based on in-orbit calibration
using Cr and Cu lines from a
modulated X-ray source and Mn from the $^{55}$Fe source. 
To assess the energy-scale correction for individual observations,
a dedicated calibration pixel—positioned outside the aperture
and continuously exposed to a collimated $^{55}$Fe source—was used. 
Gain reconstruction for this pixel relied solely on data from 
fiducial intervals applied to the
main array, while the Mn K$\alpha$ 5.9 keV line was
fitted during non-fiducial times
to emulate a celestial observation \citep{XRISM2024_N132D}.
The largest residual line offset
was found to be 0.12 eV in
OBSID 000149000, which should be
added in quadrature to the 0.3 eV systematic uncertainty. 

For each observation, spectra 
were extracted from the full array, excluding pixel 27,
which exhibited anomalous gain
shifts not aligned with the timing
of the $^{55}$Fe fiducial exposures. 
Pixel 11 also occasionally displays similar abrupt scale variations,
therefore, excluded in our analysis
\citep{A2029_Sarkar}.
Our analysis includes only high-resolution primary (Hp; \texttt{ITYPE=0}) events,
which comprise more than 99\% of the 
2--10 keV event population in each observation.
Low-resolution secondary (Ls; \texttt{ITYPE=4}) events, 
typically associated with
instrumental effects at these
low count rates, were excluded.
For background, 
a non-X-ray background (NXB) 
spectrum was generated for each observation using the
{\tt rslnxbgen} tool.
The NXB was constructed from
the Resolve night-Earth database, 
applying identical screening
criteria as for the source data. 
Weighting was applied according to 
the geomagnetic cut-off rigidity 
distribution sampled during the 
respective observation.

\begin{figure*}
    \centering
    \begin{tabular}{cc}
 \includegraphics[width=0.5\textwidth]{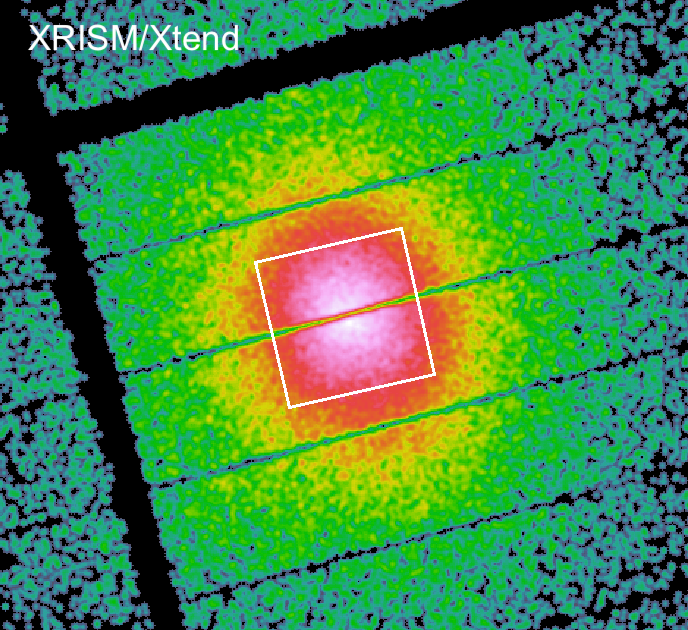} & \includegraphics[width=0.46\textwidth]{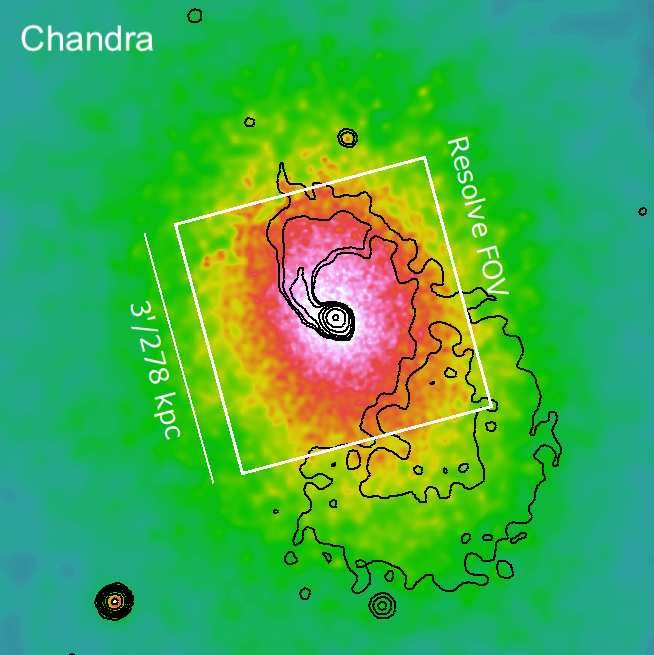}\\
  \end{tabular}
  \begin{tabular}{c}
\includegraphics[width=0.45\textwidth]{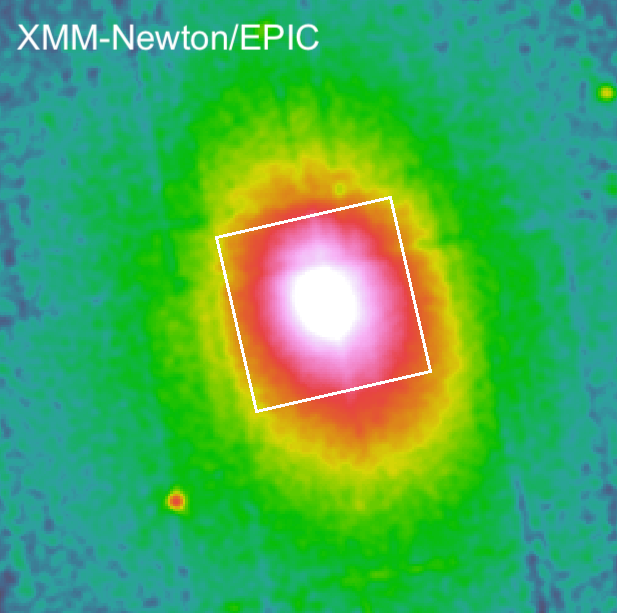}\\
  \end{tabular}
    \vspace{0.1in}
    \caption{Top-Left: XRISM/Xtend 
    count image of A2029 in the 
    0.6--10 keV energy band. 
Top-Right: Exposure-corrected,
background-subtracted Chandra
ACIS-I image of A2029 in the 
0.5--7 keV energy band.
Black surface brightness 
contours highlight the
cold-gas sloshing spiral 
in the cluster center.
Bottom: Exposure corrected,
QPB-subtracted, merged 
XMM-Newton EPIC
image of A2029. 
White boxes in all three
panels represent the 
XRISM/Resolve field of view 
(FOVs) used to extract spectra from all three instruments.}
\label{fig:chandra_image}
\end{figure*}

Instrumental responses were 
generated for each observation 
using {\tt rslmkrmf} with the most
up-to-date calibration database 
(CalDB v20250915). 
The redistribution matrix file 
(RMF) was scaled by the fraction of high-resolution primary (Hp)
events in the 2--10 keV
band—excluding low-resolution
secondary (Ls) events—to properly account for the exclusion 
of medium- and low-resolution events and ensure accurate
flux normalization. 
Ancillary response files (ARFs) 
were created using {\tt xaarfgen},
with an exposure-corrected, 
full-resolution Chandra image
in the 2–8 keV band serving as the input source model,
as shown in Figure \ref{fig:chandra_image} (right).

\subsection{XRISM/Xtend}\label{sec:xtnd_data_reduction}
The data reduction procedure for
Xtend follows the methodology
described in \citet{A2029_Sarkar}.
Briefly, the first step involved identifying and 
removing flickering pixels from the event file.
These pixels, characterized by 
abnormally high counts due to 
interactions with high-energy
cosmic rays and charged particles,
were filtered out by running the
{\tt searchflickpix} tool twice.
This approach was necessary because 
the updated version of {\tt xtdflagpix}, 
which incorporates flickering pixel removal, 
had not yet been released as part of HEASoft v6.34. 
Following this, the calibration 
sources located on either side of the Xtend field of view 
(FoV) were masked.
The event file, 
now cleaned of flickering pixels
and with calibration sources masked, 
was used to extract an image in 
the 0.6–10 keV band, as shown 
in Figure \ref{fig:chandra_image} (left). 
A spectrum was extracted from a rectangular region corresponding
to the Resolve field of view, 
also indicated in Figure 
\ref{fig:chandra_image} (left).

The response matrix file (RMF)
was generated using {\tt xtdrmf}, 
and two ancillary response files 
(ARFs) were created with {\tt xaarfgen}. 
One ARF was produced using the 
IMAGE mode, which incorporates 
a full-resolution Chandra image
as the source input;
the second used the FLATCIRCLE 
mode with a fixed radius of $15\arcmin$.
The IMAGE-mode ARF was applied 
to model the source spectra,
while the FLATCIRCLE-mode ARF was 
used to model the sky background
(see Paper III for more details; \citealt{A2029_Sarkar}).
The non-X-ray background was 
modeled using a power-law component
to represent the quiescent background, 
along with several Gaussian lines, as 
also discussed in Paper III (see 
Section 2.3).

\subsection{XMM-Newton/EPIC}
We incorporated archival 
XMM-Newton EPIC data in this
analysis, with the full list
of observation IDs 
provided in Table \ref{tab:obs_log}. 
Data reduction was carried out 
using the XMM-Newton Extended 
Source Analysis Software
(XMM--ESAS\footnote{\url{https://heasarc.gsfc.nasa.gov/docs/xmm/abc/}}) and 
associated procedures for processing EPIC 
(European Photon Imaging Camera) observations.
Event files were first calibrated 
and filtered with the standard
XMM-ESAS tasks {\tt epchain}, 
{\tt emchain}, {\tt mos-filter}, 
and {\tt pn-filter}, 
applying the most recent Current Calibration Files (CCF) database.
Soft-proton flares were removed using high-energy (2.5–12 keV) light 
curves derived from the unexposed detector corners.
From the cleaned event lists, 
images were created in the 
0.7–1.2 keV band for point-source detection.
Corresponding exposure maps were generated for each detector to 
correct for chip gaps and mirror vignetting. 
Point sources were identified
with the automated XMM-ESAS task
{\tt cheese-bands} and 
excluded from subsequent analysis. 
Quiescent particle background (QPB) images were then constructed
from filter-wheel-closed data
using the {\tt mos-back} and 
{\tt pn-back} tools 
\citep{2008A&A...478..615S}. 
The final exposure-corrected, QPB-subtracted image of A2029 is 
shown in Figure 
\ref{fig:chandra_image} (bottom).

For each observation, MOS1, MOS2, and
pn spectra were extracted from
a $3'\times3'$ region matching
Resolve FoV (see Figure 
\ref{fig:chandra_image}), using
ESAS task {\tt evselect}.
Redistribution matrix files (RMF)
were produced using {\tt rmfgen},
and ancillary response files were
generated with {\tt arfgen} task.
The QBP spectra were constructed
using {\tt mos-back} for MOS1 and MOS2
and {\tt pn-back} for the
pn instrument.
{ The hard particle
background was modeled following \citet{2015A&A...575A..37M}.}

\subsection{XMM-Newton/RGS}
XMM-Newton/RGS data reduction was
carried out with the same ESAS package
as that of EPIC.
All RGS data sets were reprocessed using {\tt rgsproc}, 
following the standard procedures recommended by the SAS team. 
Periods contaminated by solar flares were filtered using {\tt rgsfilter},
by
selecting background-quiescent intervals in the lightcurves of the
RGS 1 and 2 from CCD number 9. 
A count rate
threshold of $\sim$0.2 c/s
was applied, and the resulting 
clean exposure times are listed in 
Table \ref{tab:obs_log}.

First- and second-order RGS spectra
were extracted from a cross-dispersion 
region of width $0.8\arcmin$,
centered on the source coordinates 
(RA = $227.7328^\circ$, 
Dec = $+5.7449^\circ$).
Background spectra were constructed 
from photons lying beyond the 
98\% source point-spread function boundary;
we verified that these regions
never overlapped with the bright source. 
The {\tt rgsproc} task 
simultaneously generated the 
associated response files for both RGS1 and RGS2. 
In this study, we only considered the
first-order spectra from
both RGS instruments.

\subsection{Chandra}

We also analyzed three Chandra observations from 2000 to 2004, with the Obs ID 891, 4977 and 6101,
as reported in Table \ref{tab:obs_log}. 
There were deep Chandra ACIS-I observations in 2022 - 2023 for the cluster, with a total clean exposure time of 404.3 ks. However, the newer data require additional calibration and are in fact not much deeper than the old data because of the degradation of Chandra's response at the soft X-rays. At the 0.5 -- 7 keV band, the 2022 -- 2023 data are only $\sim$ 1.33 times deeper than the old data, measured within 10$''$ radius from the cluster center. Therefore, our studies focus on the 2000 -- 2004 data.

We reduced the Chandra ACIS observations with the Chandra Interactive Analysis of Observation (CIAO; version 4.17) and calibration database (CALDB; version 4.12.0). 
Standard ACIS data reduction was followed, which includes running 
{\tt chandra\_repro}, data mode filtering, and removal of background flares. 
Updated type 2 event 
files with clean exposure were then generated. The total clean exposure time is 106.5 ks.
We extracted the Chandra spectra within 1.5$'$ from the center and fit three spectra simultaneously. 
The background is taken from the blank sky background, with the 9.5 - 12 keV flux normalized by the values during the A2029 observations. 
All the setup for the spectral analysis is the same as for the other data.

\section{Spectral fitting}\label{sec:spectral_fitting}
We utilized the {\tt XSPEC} v12.15.0 fitting package to
perform the spectral analysis
\citep{1996ASPC..101...17A}.
All spectra were fitted 
by minimizing the C-statistics 
\citep{1979ApJ...228..939C}
to estimate the model parameters and their uncertainty range.
We assumed that the ICM 
is in collisional ionization equilibrium (CIE)
state and therefore can be modeled using the
AtomDB (v3.1.0 
\footnote{\url{http://www.atomdb.org}}).
The abundances were calculated
from all the transitions and ions of a given element, and are
scaled to the proto-solar values
of \citet{Lodders09}. 
Several previous studies have 
shown that fitting a multi-phase plasma with a single-temperature model
can lead to a systematic underestimation of the Fe
abundance \citep[e.g.,][]{2021Univ....7..208G,2022MNRAS.516.3068S}.
In Paper~III \citep{A2029_Sarkar}, 
we demonstrated that
multi-temperature gas exists
within $3'\times3'$ Resolve 
field of view at the core of A2029.
We, therefore, adopted
a two-temperature (2T)
CIE model in our spectral analysis
to mitigate this Fe bias.

\begin{figure*}
    \centering
    \begin{tabular}{cc}
\includegraphics[width=0.5\textwidth]{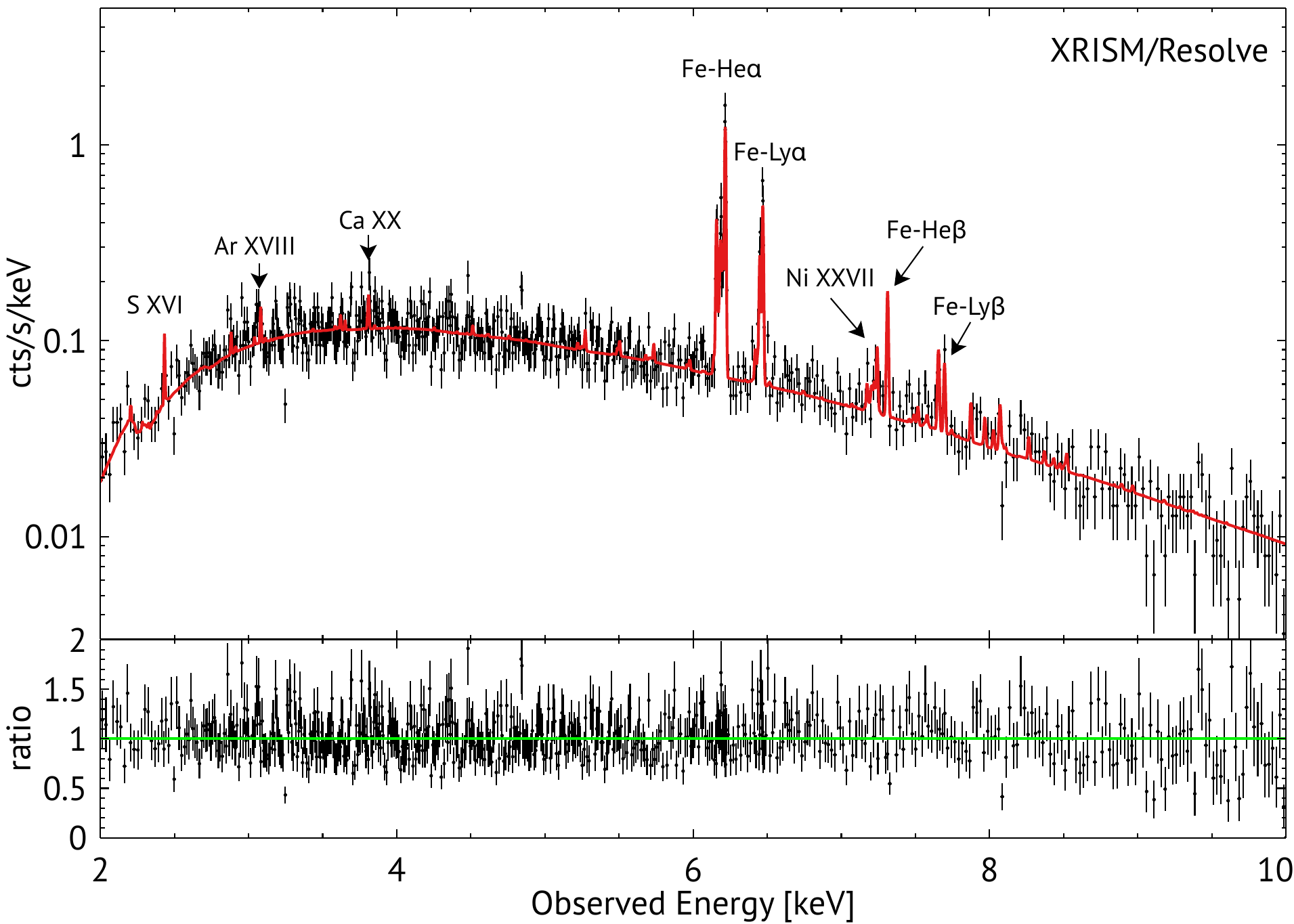} & \includegraphics[width=0.5\textwidth]{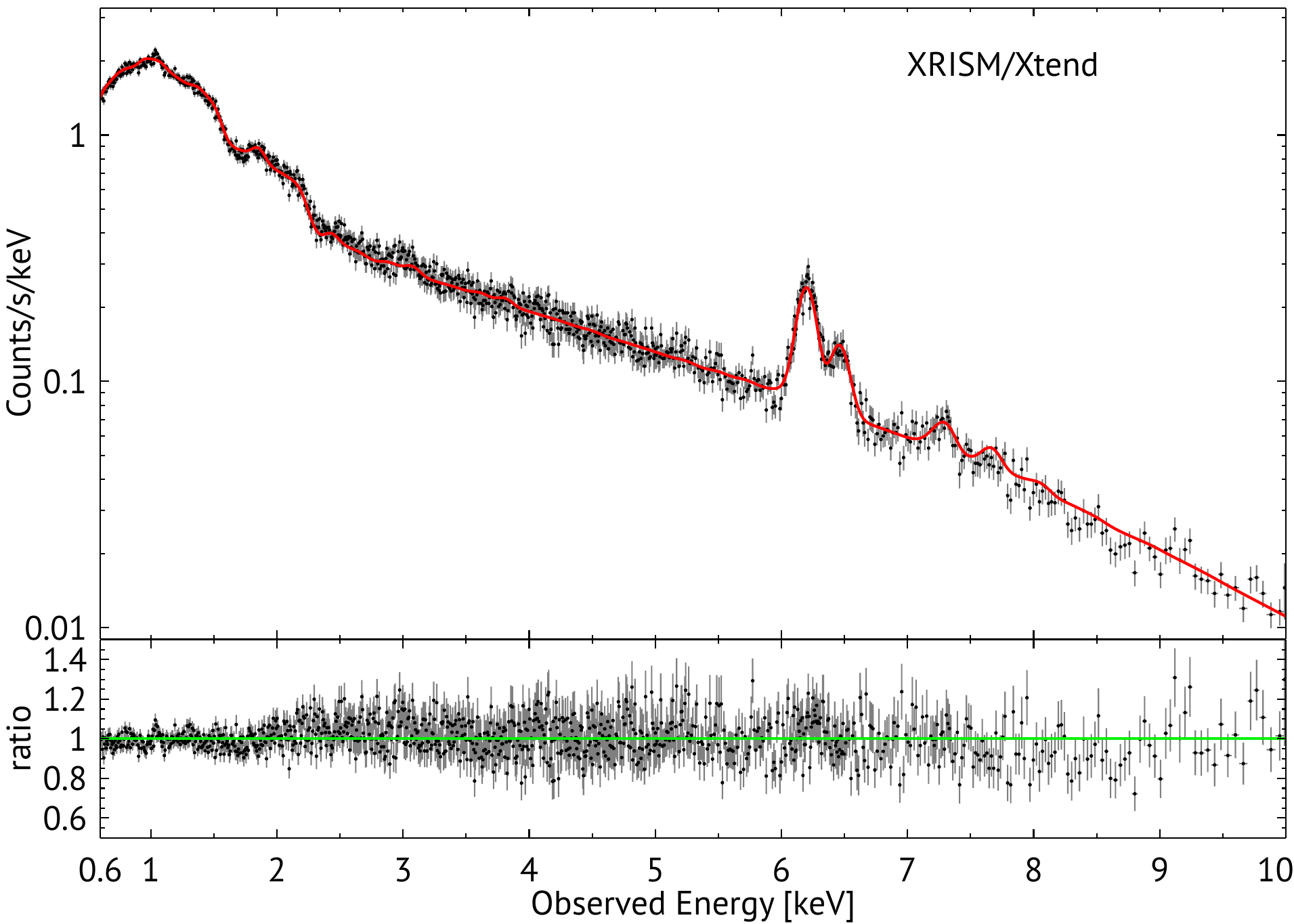}
    \end{tabular}
    \begin{tabular}{cc}
 \includegraphics[width=0.5\textwidth]{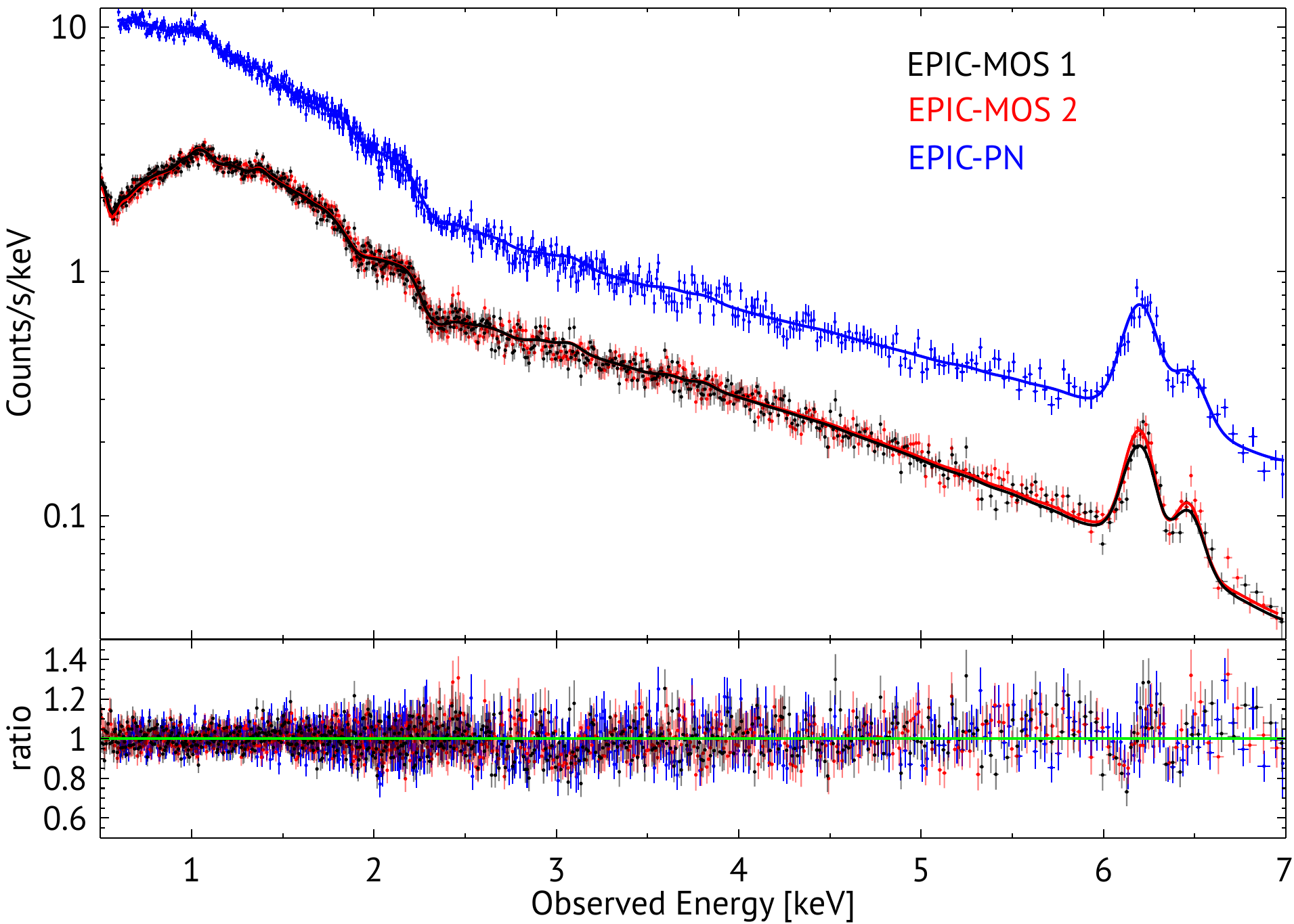} & \includegraphics[width=0.5\textwidth]{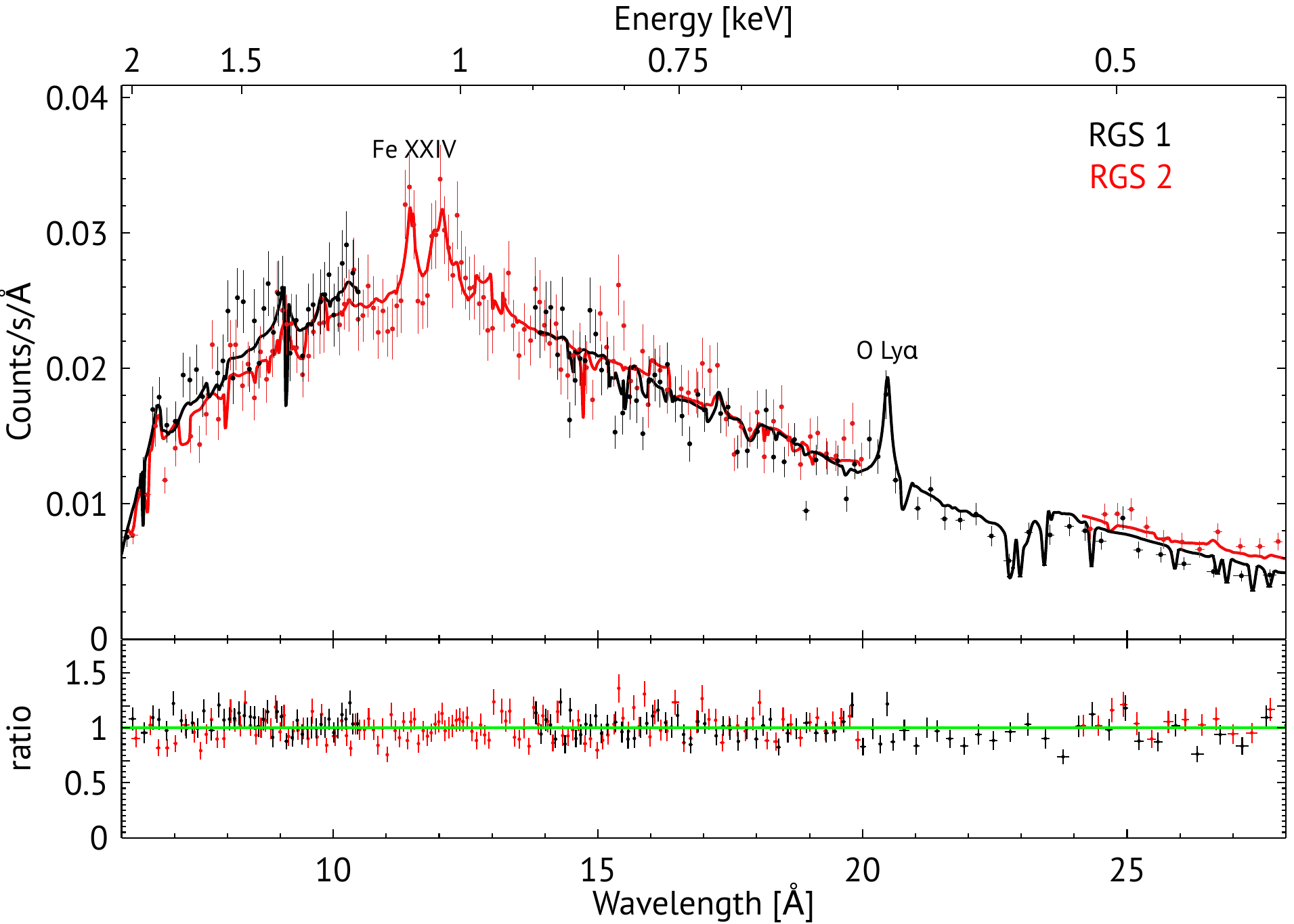}

    \end{tabular}
    \begin{tabular}{c}
\includegraphics[width=0.5\textwidth]{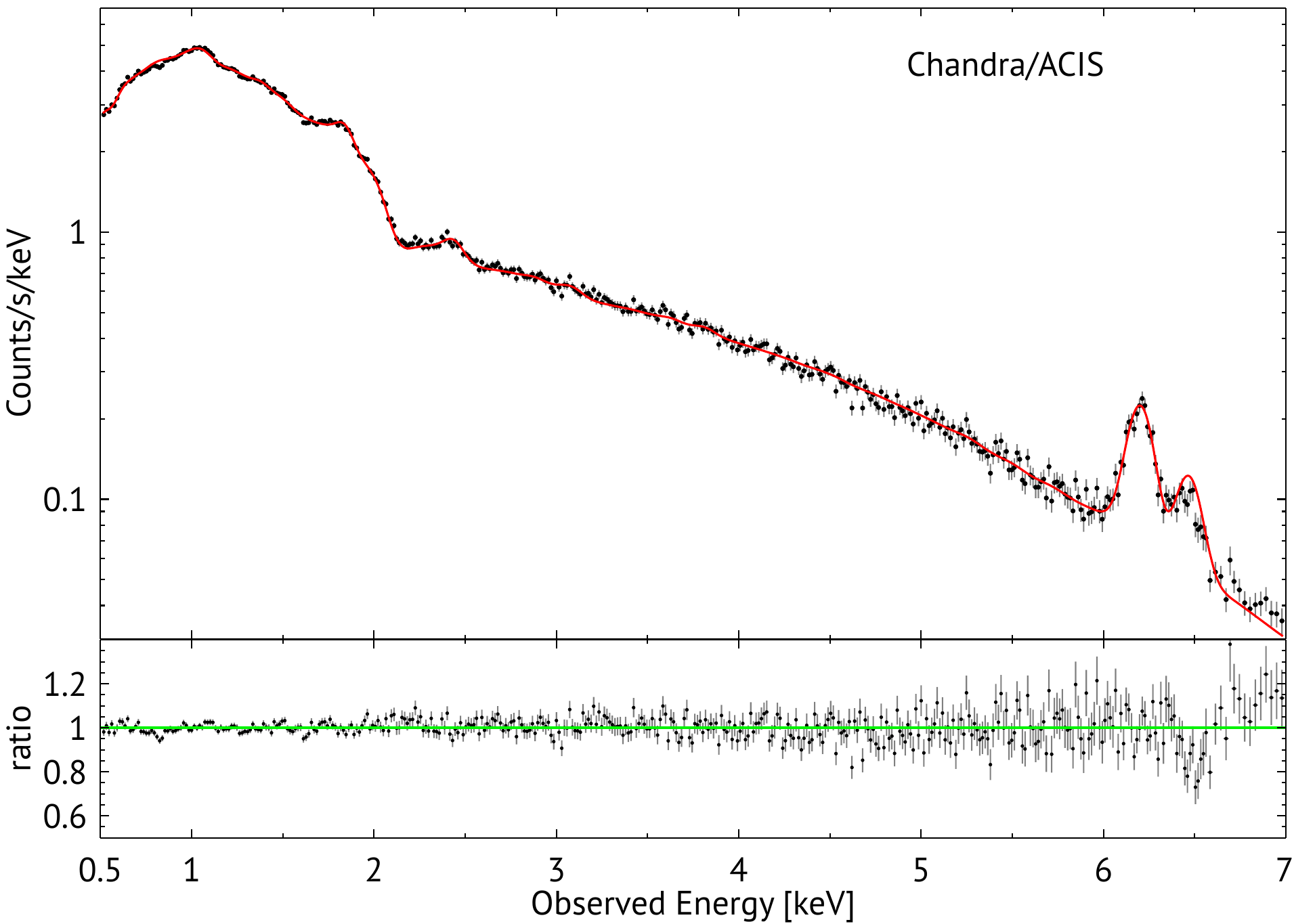}
    \end{tabular}
    \vspace{0.1in}
    \caption{{\it Top-Left:} XRISM/Resolve full-array spectrum of A2029
    (except pixels 12 and 27) from the central region, shown along with the best-fit model (red:total, blue:NXB).
    {\it Top-Right:} XRISM/Xtend spectrum
    from the central $3'\times3'$  
    region (matching Resolve FoV)
    of A2029 
    with best-fit model (red).
    {\it Middle-Left:} EPIC/MOS1 (black), EPIC/MOS2 (red),
    and EPIC/pn (blue) spectra 
    from the similar region used for
    Xtend, with best-fit models.
    {\it Middle-Right:} RGS1 (black) and RGS2 (red) first-order spectra  extracted from
    a cross-dispersion region of 
    $0.8'$, with best-fit models.
    {\it Bottom:} Chandra/ACIS spectrum
    of A2029 from the central
    $3'\times3'$ region, 
    with the best-fit model.
    For each instrument, the spectra from multiple observations have been combined to plot a single spectrum.
    In all panels, the lower sub-panels show the ratio between the observed data and the best-fit total models.
    }
    \label{fig:best_spec}
\end{figure*}

For fitting the Resolve spectra, 
we followed the same procedure as 
described in 
\citep{A2029_Sarkar}. 
The spectra were modeled with two
{\tt BVAPEC} components. 
The temperatures and normalizations
of both components were allowed to
vary, while individual elemental 
abundances, redshift, $\sigma_v$ (the line-of-sight broadening of the lines)
were varied jointly between two
components. 
All parameters were tied
between two OBS IDs.
We also modeled instrumental 
background and sky-background (such as
CXB and galactic halo) simultaneously
with the source model. 
We refer readers to \citep{A2029_Sarkar} for 
more details on individual background
models. 
Due to the loss of spectral sensitivity
below 2 keV band, we fitted 
Resolve spectra within 2--10 keV
energy band. 
This choice of spectral band
does not allow to
constrain the abundances of O, Ne, Mg, and
Si.
We, therefore, tied
these elements to Fe while fitting.
Figure \ref{fig:best_spec} (top-left)
shows the Resolve spectra of A2029 
with the best-fit model.

Xtend spectral fitting 
was performed in the 0.6--10 keV
energy band, as recommended 
in the quick start 
guide{\footnote{\url{https://heasarc.gsfc.nasa.gov/docs/xrism/analysis/quickstart}}}. 
We fitted Xtend spectra with two
{\tt VAPEC} components for
ICM emission simultaneously with
the NXB and sky background models, following 
the same strategy as 
\citet{A2029_Sarkar}. 
Figure 
\ref{fig:best_spec} (top-right)
panel shows the Xtend spectra 
for the A2029 core alongwith 
the best-fit model.
The temperature and normalizations
of both components were free
to vary however tied between 
OBS IDs. 
The elemental
abundances 
of O, Ne, Si, S, Fe, and Ni
were varied jointly
between two {\tt VAPEC} components.
The redshift parameter was fixed
to the best-fit value 
($z=$0.07776) obtained from
Resolve spectral fitting.

We analyzed archival XMM-Newton
EPIC and RGS observations 
to complement XRISM
measurements. 
{ For the EPIC, the MOS and pn spectra 
from all OBS IDs were fitted simultaneously 
using
two {\tt VAPEC} components for
the ICM emission and 
a full background model for the
sky background
(including cosmic X-ray background and galactic foreground components), similar to the
Xtend.}
The temperature of two {\tt VAPEC} components were free to vary 
however were tied between MOS and pn
as well as between all OBS IDs. 
However, normalization components
were free to vary for all components,
instruments, and OBS IDs. 
The elemental abundances of O, Ne,
Mg, Si, S, Fe, and Ni 
were varied collectively
between instruments and OBS IDs.
Previous XMM-Newton studies 
such as \citet{2015A&A...575A..30S}
reported extremely variable soft tail
in the EPIC filter wheel closed events that might considerably
aﬀect the spectra below 0.5 keV.
\citet{2016A&A...592A.157M} 
found a good compromise by
fitting the MOS spectra
within 0.5–7 keV band and 
pn spectra within 0.6–7 keV band.
We followed the same in this paper,
as presented in Figure 
\ref{fig:best_spec} (middle-left).

For this paper we only consider
the first order spectra for
XMM-Newton/RGS. 
Spectra from RGS 1 and RGS 2
instruments for all OBS IDs were
fitted simultaneously 
with 
two {\tt VAPEC} components to
model the ICM emission. 
For each spectrum, we took into account the
spectral broadening effect due to the spatial extent of the source
by using the 
{\tt rgsxsrc} multiplicative model
in XSPEC and supplying MOS images.
The temperature of the hotter component was 
free to
vary, tied between instruments
and OBS IDs.
The cooler temperature component was
linked to the hotter component as
$kT_{\rm hot}:kT_{cool}=1:0.25$, 
following \citet{2023ApJ...953..112F}.
We did not adopt any assumption regarding the emission measures of each temperature component, therefore,
left it free to vary.
The spectral fitting was performed
in the 0.2--2.0 KeV energy band.
Due the limited spectral coverage
of RGS, 
the elemental abundances 
of O, Ne, Mg, Si, and Fe 
were varied freely and 
other elements collectively with Fe,
however all were tied between two 
{\tt VAPEC} components.
The redshift parameters
were fixed to the best-fit value
obtained from Resolve spectral 
fitting.
Figure \ref{fig:best_spec} 
(middle-right) shows the 
RGS 1 and RGS 2
spectra summed over all OBS IDs
along with the best-fit models.

For the Chandra ACIS, we followed the
similar fitting strategy as EPIC and
Xtend. We modeled ACIS spectra
with two absorbed {\tt VAPEC} components
to model the source emission. 
Figure \ref{fig:best_spec} (bottom)
shows the ACIS spectra from the central
region of A2029 along with the best-fit
model. 
The elemental abundances of O, Ne,
Mg, Si, S, Fe, and Ni were free to
vary but tied between two temperature
components. 
The best-fit parameters are listed
in Table \ref{tab:best_params}.

For each spectral fitting described 
above, to account for the Galactic 
neutral column density along
the line of sight,
we applied the multiplicative 
{\tt phabs} absorption model 
in {\tt XSPEC} along with CIE components 
\citep{2000ApJ...542..914W}.
The absorption cross sections 
were adopted from 
\citet{1996ApJ...465..487V}.
We fixed to the galactic value,
i.e., $N_{\rm H} = 3\times10^{20}$ cm$^{-2}$ 
\citep{2016A&A...594A.116H}.

\section{Results}
\subsection{Temperature measurement}\label{sec:elem_abun}
The best-fit temperatures from 
Resolve spectral fitting indicate the presence of two components:
a cooler phase at $4.81^{+0.62}_{-0.42}$ keV and a hotter phase 
at $8.97^{+2.77}_{-0.65}$ keV. 
The corresponding redshift 
($z = 0.07776^{+0.00004}_{-0.00003}$) and 
velocity dispersion 
($\sigma_v = 155^{+11}_{-11}$ km s$^{-1}$) are consistent
with previous XRISM 
studies \citep[e.g.,][]{A2029_Sarkar,A2029_naomi,A2029_eric}.
The best-fit Xtend 
temperatures—$5.85^{+0.52}_{-0.47}$ keV 
and $10.66^{+0.64}_{-0.65}$ keV—are 
also in agreement with those
measured by Resolve and earlier
XRISM studies.
Similarly,
the cooler gas temperature measured
by EPIC (4.60$^{+0.11}_{-0.15}$ keV) 
and hotter gas temperature measured by 
RGS (4.40$^{+0.11}_{-0.10}$ keV)
are consistent with the cooler gas 
found using Resolve and Xtend.
The hotter temperature component
from EPIC 
($10.62^{+0.40}_{-0.39}$ keV)
is likewise consistent with the
hotter components measured by 
Resolve and Xtend.
We note that the RGS spectral band is restricted to
0.2--2 keV.
Therefore, it is not particularly
sensitive to the hotter gas,
which is better constrained by
Resolve, Xtend, and EPIC. 
From Chandra spectral fitting,
we obtain a cooler gas temperature 
of $5.32^{+0.37}_{-0.21}$ keV,
in agreement with the 
gas temperatures found by
Resolve, Xtend, EPIC, and RGS.
However, we find that the hotter
gas temperature measured with 
Chandra ($12.9^{+1.47}_{-0.39}$ keV)
is marginally higher than
those measured by the other 
instruments. This is consistent with the trend found by \cite{2015A&A...575A..30S} in a comparison of cluster temperatures measured by XMM-Newton and Chandra.

\subsection{Elemental abundances}
\begin{figure*}
    \centering
    \begin{tabular}{c}
       \includegraphics[width=0.7\textwidth]{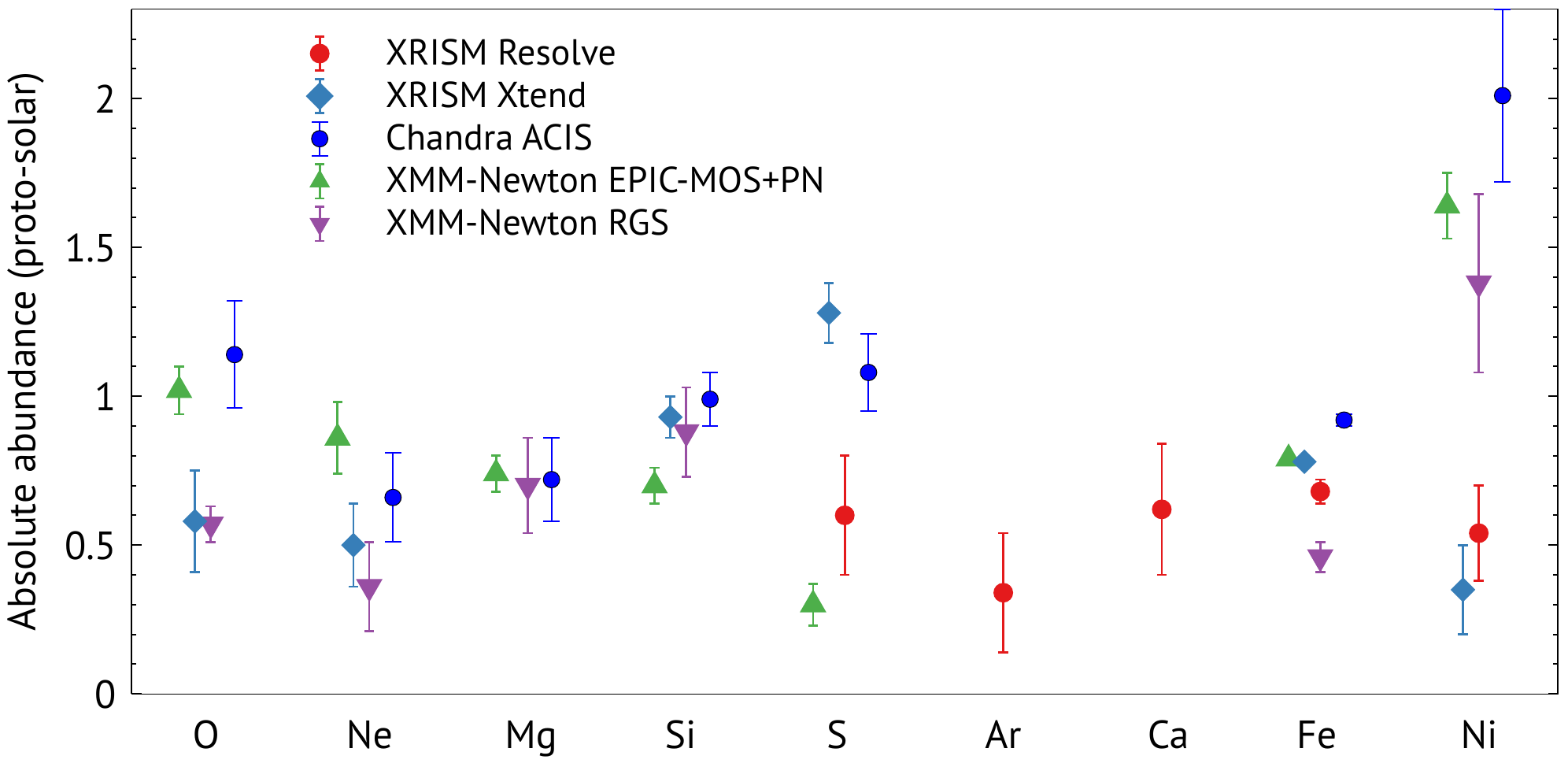}\\  \includegraphics[width=0.7\textwidth]{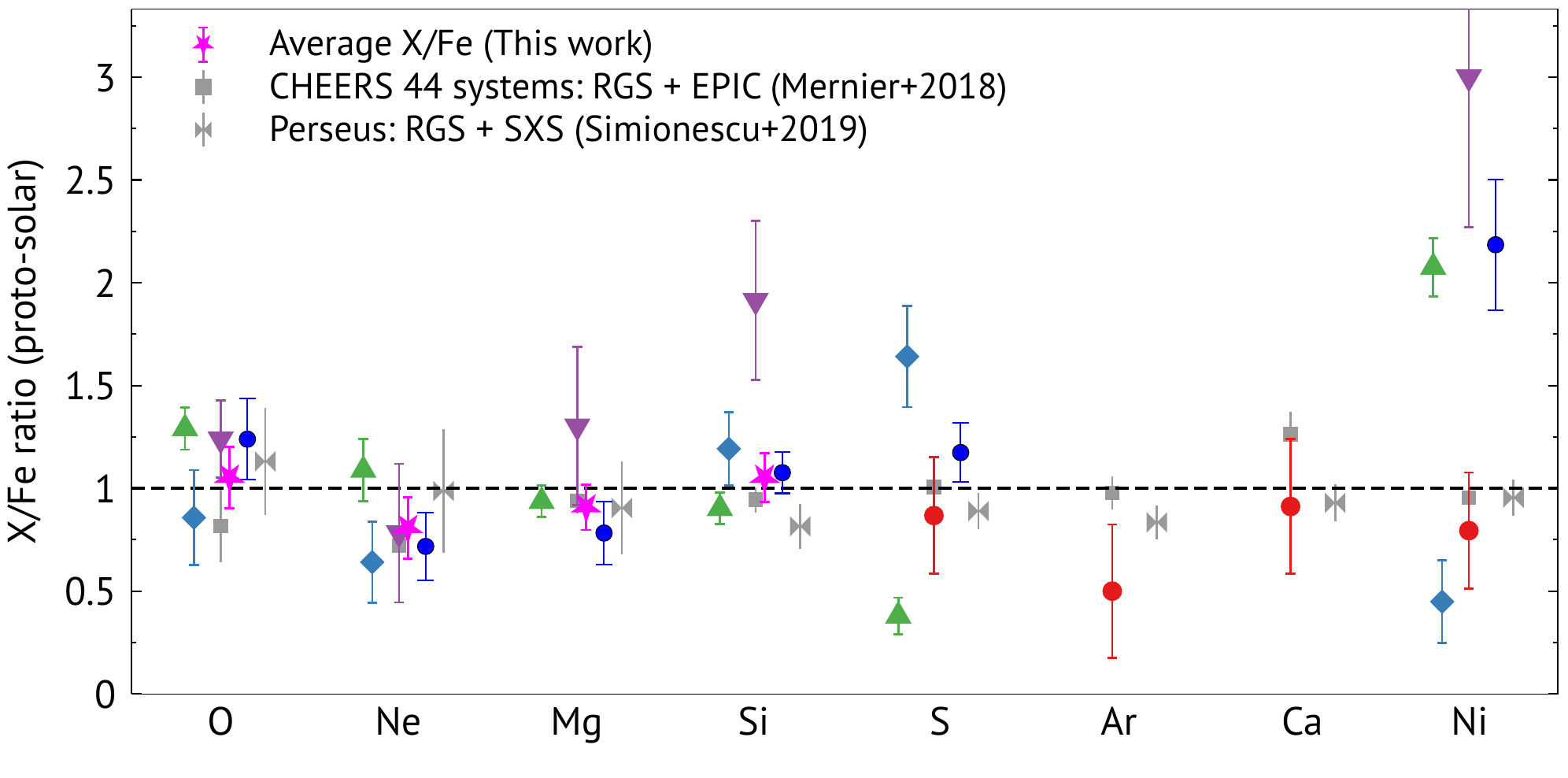}\\
    \end{tabular}
    \caption{Top: Absolute elemental
    abundance of different light and heavy
    elements 
    in the central region of A2029 measured using XRISM (Resolve+Xtend), XMM-Newton (MOS+pn, RGS), and Chandra (ACIS). 
    Bottom: Abundance ratios of different elements with respective to Fe (X/Fe). Grey boxes and tiehorizontals represents the 
    X/Fe for CHEERS samples of 
    clusters/groups \citep{2018MNRAS.478L.116M} and Perseus cluster \citep{2019MNRAS.483.1701S}.}
    \label{fig:central_abun_x_fe}
\end{figure*}

Figure \ref{fig:central_abun_x_fe} (left) shows the best-fit abundances 
of O, Ne, Mg, Si, S, Ar, Ca, Fe, and
Ni within the central $3'$ of A2029. 
The corresponding numerical values
are listed in Table \ref{tab:best_params}. 
Thanks to its exceptional 
spectral resolution, 
Resolve provides not only precise
but also \textit{accurate} measurements of abundances from emission lines due to its ability to more cleanly separate the lines from the underlying continuum compared to CCD instruments. 
The best-fit abundances measured
with Resolve are:
S = $0.60^{+0.18}_{-0.17}\ Z_{\odot}$,
Ar = $0.34^{+0.21}_{-0.20}\ Z_{\odot}$,
Ca = $0.62^{+0.22}_{-0.20}\ Z_{\odot}$,
Fe = $0.68^{+0.04}_{-0.04}\ Z_{\odot}$, and
Ni = $0.54^{+0.17}_{-0.15}\ Z_{\odot}$.

\begin{table*}
\caption{Best-fit parameters obtained by fitting spectra from the central region of A2029.\label{tab:best_params}}   
\begin{center}
\setlength{\tabcolsep}{6pt}
\begin{tabular}{lccccc}
Parameter & {Resolve} & {Xtend} & EPIC & RGS & Chandra\\ 
\hline
\hline
$kT_{1}$ (keV) & 4.81$_{-0.42}^{+0.62}$ & 5.85$_{-0.47}^{+0.52}$ & 4.66$_{-0.15}^{+0.11}$ & 4.40$_{-0.10}^{+0.11}$ & 5.32$_{-0.21}^{+0.37}$\\
$kT_{2}$ (keV) & 8.97$_{-0.65}^{+2.77}$ & 10.66$_{-0.65}^{+0.64}$ & 10.62$_{-0.39}^{+0.40}$ & 1.10\tablenotemark{\footnotesize d} & 12.9$_{-1.45}^{+1.47}$\\
O & $-$ & 0.58$_{-0.17}^{+0.16}$ & 1.03$_{-0.08}^{+0.08}$ & 0.57$_{-0.08}^{+0.06}$ & 1.14$_{-0.18}^{+0.18}$\\
Ne & $-$ & 0.50$_{-0.13}^{+0.15}$ & 0.86$_{-0.12}^{+0.12}$ & 0.36$_{-0.15}^{+0.15}$ & 0.66$_{-0.14}^{+0.16}$\\
Mg & $-$ & $<0.35$\tablenotemark{\footnotesize a} & 0.74$_{-0.06}^{+0.06}$ & 0.70$_{-0.16}^{+0.16}$ & 0.72$_{-0.13}^{+0.13}$\\
Si & $-$ & 0.93$_{-0.06}^{+0.07}$ & 0.70$_{-0.05}^{+0.05}$ & 0.88$_{-0.15}^{+0.15}$ & 0.99$_{-0.08}^{+0.08}$\\
S & 0.60$_{-0.17}^{+0.18}$ & 1.28$_{-0.09}^{+0.11}$ & 0.30$_{-0.05}^{+0.05}$ & $-$ & 1.07$_{-0.11}^{+0.12}$\\
Ar & 0.34$_{-0.20}^{+0.21}$ & $-$ & $-$ & $-$ & $-$\\
Ca & 0.62$_{-0.20}^{+0.22}$ & $-$ & $-$ & $-$ & $-$\\
Fe & 0.68$_{-0.04}^{+0.04}$ & 0.77$_{-0.01}^{+0.01}$ & 0.79$_{-0.01}^{+0.01}$ & 0.46$_{-0.04}^{+0.04}$ & 0.90$_{-0.02}^{+0.02}$\\
Ni & 0.54$_{-0.15}^{+0.17}$ & 0.31$_{-0.10}^{+0.14}$ & 1.64$_{-0.11}^{+0.11}$ & 1.38$_{-0.30}^{+0.30}$ & 2.01$_{-0.32}^{+0.32}$\\
Redshift & 0.07776$_{-0.00003}^{+0.00004}$ & 0.07776\tablenotemark{\footnotesize b} & 0.07776\tablenotemark{\footnotesize b} & 0.07776\tablenotemark{\footnotesize b} & 0.0777\tablenotemark{\footnotesize b}\\
$\sigma_v$ (km/s) & 155$_{-11}^{+11}$ & $-$ & $-$ & $-$ & $-$\\
$N_1$ (10$^{12}$ cm$^{-5}$)\tablenotemark{\footnotesize c} & 1.85$_{-0.72}^{+0.78}$ & 2.09$_{-0.3}^{+0.4}$ & 1.85$_{-0.19}^{+0.15}$ & 2.40$_{-0.19}^{+0.15}$ & 0.04$_{-0.03}^{+0.03}$\\
$N_2$ (10$^{12}$ cm$^{-5}$)\tablenotemark{\footnotesize c} & 2.84$_{-0.57}^{+0.81}$ & 2.2$_{-0.3}^{+0.3}$ & 1.53$_{-0.15}^{+0.19}$ & $<0.003$ & 4.02$_{-0.05}^{+0.05}$\\
Energy Band (keV) & 2--10 & 0.6--10 & 0.5--7 & 0.2--2 & 0.5--7\\
C-stat/dof & 3847/3846 & 5989/5737 & 17725/16698 & 3246/2935 & 1413/1318\\ 
\hline
\end{tabular}
\vspace*{-\baselineskip}
\end{center}
\tablenotetext{a}{\footnotesize 3$\sigma$ upper-limit}
\tablenotetext{b}{\footnotesize Redshift was fixed.}
\tablenotetext{c}{\footnotesize The model normalization corresponds to a 3-arcmin diameter circular region centered on Abell 2029.}
\tablenotetext{d}{\footnotesize The temperature parameter was linked with
the hotter component as $kT_{\rm hot}:kT_{\rm cool}=1:0.25$, following \citet{2023ApJ...953..112F}.}
\end{table*}

The Fe abundances measured by Resolve, 
Xtend, and XMM-Newton/EPIC 
are marginally consistent with each other. 
However, the Fe abundance derived 
from XMM-Newton/RGS is systematically lower than those
from other instruments. 
This discrepancy is expected and
can be attributed to several factors. 
The RGS extraction region differs in angular coverage, and so this could reflect real spatial abundance variations. In addition, the RGS bandpass excludes the Fe-K complex, so the Fe abundance is solely constrained by the Fe-L lines, which are in a forest of additional spectral lines that complicate proper measurement of the continuum. 
Moreover, for extended sources
like A2029, instrumental 
line broadening further degrades the RGS ability to distinguish line emission from continuum, 
leading to larger systematic uncertainties. 
Similar differences between 
RGS and EPIC Fe measurements have been reported in previous studies, 
such as \citet{2017A&A...607A..98D} and \citet{2016A&A...592A.157M} 
based on the CHEERS sample.
In contrast, Chandra yields a significantly higher
Fe abundance of $0.90 \pm 0.02$ solar. 
{ A higher Fe abundance
in Perseus core measured with 
ACIS-S3, compared to EPIC, was also reported by \citet{2009A&A...493...13M}. 
They attributed the difference to cross-calibration uncertainties in the 1–5 keV effective area rather than to a real variation in abundance.
However, we adopt Chandra  measurements in this analysis because the X/Fe ratios are consistent with those from other instruments.}

Absolute elemental abundances are 
often more sensitive to systematic uncertainties than
relative abundances. 
It is, therefore, generally
more reliable to examine abundance 
ratios relative to Fe (X/Fe).
Figure \ref{fig:central_abun_x_fe} 
(right) presents the derived 
X/Fe abundance ratios for all 
measured elements at the core of A2029.
The O/Fe ratios measured by RGS, 
EPIC, and Chandra are consistent with each other
but are marginally higher 
than that derived from Xtend. 
{ The Ne/Fe
ratios from Xtend, EPIC, RGS, and Chandra agree within their
$1\sigma$ uncertainties.}
We note that while EPIC 
provide robust constraints 
on many elemental abundances,
measuring O and Ne remains 
challenging. 
The $\oviii$ emission line lies near $\sim$0.6 keV.
{ The EPIC spectral fitting were restricted to $\geq$ 0.5 keV for
MOS and $\geq$ 0.6 keV
for pn,
close to the instrumental oxygen absorption edge, 
where calibration uncertainties
can be significant.} 
Additionally, the Ne K-shell
lines overlap with the Fe-L complex, 
making Ne abundance measurements 
highly sensitive to assumptions
about the ICM temperature
structure.

The Mg/Fe abundance ratios
are consistent between the 
EPIC and Chandra measurements.
Although RGS yields an absolute 
Mg abundance that agrees with 
EPIC, 
its Mg/Fe ratio is marginally higher, 
driven by the systematically lower Fe abundance derived from RGS. 
For Xtend, we were unable to
obtain a well-constrained 
Mg abundance; 
instead, we derive an 
3$\sigma$ upper
limit of $<$0.35 solar,
corresponding to Mg/Fe $<$ 0.46.
Si/Fe abundance ratios can be
measured with relatively high precision using CCD instruments. 
As shown in Figure \ref{fig:central_abun_x_fe},
the Si/Fe ratios derived from Xtend, EPIC, and Chandra are consistent.
In contrast, RGS shows a systematically
higher value. 
The $\sixiii$ He-like triplet 
(at 1.865 keV) and 
$\sixiv$ Ly$\alpha$ line 
(at $\sim$2.0 keV) 
lie near the 
edge of the RGS bandpass, 
where the instrument’s effective 
area declines rapidly. 
As a result, the Si/Fe ratios
measured by RGS are less reliable
for extended sources like A2029.

For the S/Fe, Ar/Fe, Ca/Fe, and
Ni/Fe abundance ratios, 
we primarily leverage Resolve’s superior spectral resolution
and sensitivity, 
which offer significant advantages
over CCD-based instruments.
In addition to Fe He-like and H-like 
lines, $\sxv$ He-like triplet at 
$\sim$2.46 keV, $\arxvii$ at $\sim$3.12 keV, $\caxix$ at $\sim$3.90 keV, 
and $\nixxvii$ at $\sim$7.8 keV 
are all clearly detected in
the core of A2029 with Resolve.
This capability allows for
precise abundance measurements of
these elements, 
which are otherwise blended
with the continuum or neighboring
lines in CCD spectra,
where limited resolution and 
high instrumental background 
introduce substantial systematic uncertainties.
Resolve clearly
resolves the Ni–K$\alpha$ ($w$)
line in the A2029 core
(Figure \ref{fig:best_spec}),
resulting in a subsolar Ni/Fe 
abundance ratio.
{ Since, EPIC and Chandra spectral fitting were restricted to 7 keV and
RGS can not resolve any $\nixxvi$
lines,
the Ni abundance measured
using these instruments are not reliable.} 
Xtend, because of its orbit,
benefits from a significantly lower instrumental background
compared to EPIC, RGS, and Chandra, 
enabling a more reliable Ni
abundance measurement that 
is in good agreement with Resolve.

Taking into account the strengths 
and limitations of each instrument, 
we adopt the following approach
for the remainder of the analysis.
For O/Fe, we use the average of 
RGS and Xtend measurements. 
{ For Ne/Fe, we adopt the average
over Xtend, EPIC, RGS, and Chandra
as these also show mutual agreement.} 
The Mg/Fe ratio is averaged over
EPIC and Chandra, 
while Si/Fe is averaged 
across Xtend, EPIC, and Chandra, 
excluding RGS due to its limited sensitivity to Si lines 
in extended sources.
Finally, for S/Fe, Ar/Fe, Ca/Fe, 
and Ni/Fe, we adopt the values 
measured with Resolve.
The adopted average X/Fe
abundance ratios at the center
of A2029 are shown in Figure 
\ref{fig:central_abun_x_fe} (right), marked in magenta.

\subsection{Systematic uncertainty}
We examine the measured chemical abundances in the core of A2029 for 
possible systematic biases. 
Because our analysis combines observations from multiple telescopes,
the primary source of bias arises from 
cross-instrument comparisons. 
As shown in Figure \ref{fig:central_abun_x_fe}, 
most X/Fe measurements are consistent across the different instruments. 
Outlier values were excluded when constructing the average abundance pattern; for example, 
we omitted the RGS measurement of Si/Fe.
To further reduce potential biases,
we re-analyzed the archival spectra using the same fitting procedure applied to the XRISM data,
rather than adopting chemical abundances 
reported in the literature.

Any systematic uncertainties in the continuum modeling can 
significantly affect the derived metal abundances, because,
the strength of emission lines 
is measured relative to the local continuum. 
In each of the spectral fitting cases discussed in Section \ref{sec:spectral_fitting}, the modeled continuum includes contributions from multiple thermal components
in the CIE model, background emission (NXB + Sky), and possible power-law emission
from a central AGN. 
All components are subject to absorption by the Galactic interstellar medium (ISM).
Since our focus is limited to the 
brightest central region of A2029,
both AGN and background components do not play a major role and 
thermal emission from ICM dominates in
all cases \citep{2017A&A...607A..98D,A2029_Sarkar}.

The Galactic absorption column is typically
measured via radio surveys and expressed as the atomic hydrogen column density, 
$N_{\mathrm{H}}$ \citep{2005A&A...440..775K,2013MNRAS.431..394W}. 
In dense regions of the Galaxy,
however, molecular hydrogen and dust can also contribute to the total absorption. 
As a result, X-ray-derived $N_{\mathrm{H}}$ values may differ from those inferred from radio 
or multiwavelength observations, potentially affecting abundance measurements 
\citep[e.g.,][]{2016A&A...592A.157M,2017A&A...607A..98D}.
To assess potential systematic uncertainties 
due to Galactic absorption, we varied the $N_{\mathrm{H}}$ value by $\pm10\%$ from the 
adopted baseline of $N_{\mathrm{H}} =
3 \times 10^{20}$ cm$^{-2}$ \citep{2016A&A...594A.116H}. 
Since the XRISM/Resolve energy band is restricted to $>2$ keV,
it is less sensitive to Galactic absorption 
effects than other instruments such as RGS. 
However,
these variations in $N_{\mathrm{H}}$ had no significant impact on
the best-fit parameters, as reported in Table
\ref{tab:best_params}.

\section{Discussion}
\subsection{Cluster core enrichment}

In this work, we measure the chemical abundances of several elements in the core of A2029 using XRISM (Resolve and Xtend). The exceptional spectral resolution of Resolve allows us to resolve individual emission lines, enabling robust abundance measurements of elements such as S, Ar, Ca, Fe, and Ni. However, because the closed gate valve prevents Resolve from providing spectral sensitivity below 2.0 keV,
we supplement the XRISM data with archival XMM-Newton (EPIC and RGS) and Chandra (ACIS) observations. These instruments provide coverage below 2.0 keV, allowing reliable measurements of lighter elements including O, Ne, Mg, and Si. Together, this joint analysis demonstrates the power of 
combining XRISM with existing observatories, yielding robust abundance constraints for nine elements in total.
Among these, Ar, Ca, Fe, and Ni predominantly originate from Type Ia supernovae, while O, Ne, and Mg are mainly produced in core-collapse supernovae.

Since absolute abundances may depend on
the instrumental and other systematic
effects (e.g., temperature structures),
we focus on deriving abundance ratios relative to Fe (X/Fe). 
Figure \ref{fig:central_abun_x_fe}
presents the resulting X/Fe ratios
at A2029 core.
Previous studies by \citet{2016A&A...592A.157M}
and \citet{2017A&A...607A..98D}
reported X/Fe ratios for A2029 
within 0.2R$_{500}$ and 0.05R$_{500}$ using EPIC and RGS. 
Our independently derived 
X/Fe ratios from EPIC and RGS, such as 
O/Fe ($\sim 1.3\pm0.2$), Ne/Fe ($\sim0.8\pm0.3$), Mg/Fe ($\sim0.9\pm0.1$), Si/Fe ($\sim0.9\pm0.1$), and 
S/Fe ($\sim0.4\pm0.1$)—are consistent with those results.
The main difference arises in Ni/Fe: 
earlier CCD-based analyses
obtained significantly higher 
values than our XRISM measurements. 
This discrepancy likely stems from
the limited spectral resolution of
CCD instruments, which
prevents robust Ni/Fe constraints because the 
Ni–K lines remain unresolved. 
In addition, the high-energy band 
around the Ni–K transitions is
strongly influenced by instrumental background, 
as the cluster emission drops 
steeply at these energies. 
In contrast, Resolve clearly
resolves the Ni–K line 
(see Figure \ref{fig:best_spec}), 
allowing a precise abundance determination. 
Consistently, Xtend—unlike Chandra
and EPIC—benefits from a lower 
particle background, 
yielding robust Ni abundance.

The similarity of Fe abundance
patterns in massive clusters \citep{2017MNRAS.470.4583U} and
galaxy groups \citep{2022MNRAS.516.3068S} 
at large radii indicates a common 
early enrichment process predating gravitational collapse. This similarity, and the large mass of diffuse metals compare to the number of stars in present day cluster galaxies, suggests that cluster and group atmospheres were enriched between $3 < z < 10$ prior to the formation of clusters 
\citep[e.g.,][]{2001ApJ...557..573L,2021arXiv210504638E}.  
The supernova ejecta was subsequently transported and mixed into the protocluster gas 
since $z \sim 3$ by large-scale radio jets and galactic winds.

In contrast, the pronounced
central abundance peaks observed in massive cool-core clusters
are generally attributed to metal enrichment from the brightest
cluster galaxy (BCG). 
Most BCGs are quiescent, 
appearing red and dead, with their 
present-day star formation rates
on the order of $\sim0.1\ M_{\odot} 
{\rm yr}^{-1}$.
Cooling flow systems, however, 
exhibit elevated activity, with star formation rates reaching a 
few $M_{\odot} {\rm yr}^{-1}$, 
and in rare, extreme cases, as 
high as several hundred $M_{\odot} {\rm yr}^{-1}$ 
\citep{2012Natur.488..349M,2018ApJ...858...45M}. In these systems, enhanced light elements are expected in the central regions reflecting enhanced core collapse supernova activity.

Despite being a strong cooling flow, optical observations of A2029's central cD galaxy, IC 1101, 
have obtained only an upper limit on the  central H$\alpha$ luminosity of  $<$ 4.4 $\times$ 10$^{39}$ erg s$^{-1}$.  This upper limit implies a visible star formation rate of
$<0.03M_{\odot}$ yr$^{-1}$
\citep{2010ApJ...721.1262M,2010ApJ...719.1844H,2012NJPh...14e5023M}.
The chemical enrichment in the cluster core is therefore thought to be dominated by SNIa with long delay times, supplemented to a lesser degree  by metal-rich stellar winds.
AGN-driven uplift,
on the other hand, redistributes  
metals produced by
both SNcc and
SNIa that locked up in the stars and the gas in the BCG.
SNIa explosions are believed to dominate the majority of Fe and
Ni enrichment in clusters.

Abundances of elements heavier than S in A2029 core are consistent with those of the Perseus cluster 
near NGC 1275 \citep{2017Natur.551..478H}, which both lie near the 
Solar values.  
This is true despite the very different star formation histories in
NGC 1275 and IC 1101.  
NGC 1275 is presently forming stars at $\sim 10~\rm M_\odot ~ yr^{-1}$.  
This would lead to enhancements both in light and heavy elements due to enhanced SNcc in NGC 1275
compared to IC 1101, 
although SNcc preferentially produce lighter elements that Resolve is not sensitive to.
XRISM has revealed abundances in the Centaurus cluster which are  above those of the Perseus cluster and the solar value (Mernier et al. 2025, submitted). 

The average abundance pattern of  the elements in A2029 core,
relative to Mg provides a useful comparison to 
the stellar abundance pattern in Milky Way
from the APOGEE surveys \citep[e.g.,][]{2015ApJ...808..132H,2019ApJ...874..102W}. In the 
Galactic disk, X/Mg ratios derived from
APOGEE data shows that $\alpha$-elements
such as O, Si, S, and Ca track Mg closely,
while Fe-peak elements like Ni increase
with Mg due to the delayed contribution of
SNIa. 
The near-solar Mg/Fe ratio at A2029 center 
therefore imply a well-mixed enrichment
by both SNIa and SNcc, similar to the 
late-time regime of the Milky Way's thin 
disk. 
The sub-solar S/Mg, Ar/Mg, Ca/Mg,
and Ni/Mg ratios at A2029 center may
arise from differences in integrated SNe 
yields, where certain elements are 
produced less efficiently in SNcc or SNIa
in cluster galaxies, perhaps indicating time dependent or non-universal SN yields. 
{ These sub-solar ratios may also,
however,
be influenced by cross-instrumental 
biases.}

\subsection{SNIa and SNcc contributions in metal enrichment}

\begin{figure*}
    \centering
    \begin{tabular}{cc}
       \includegraphics[width=0.5\textwidth]{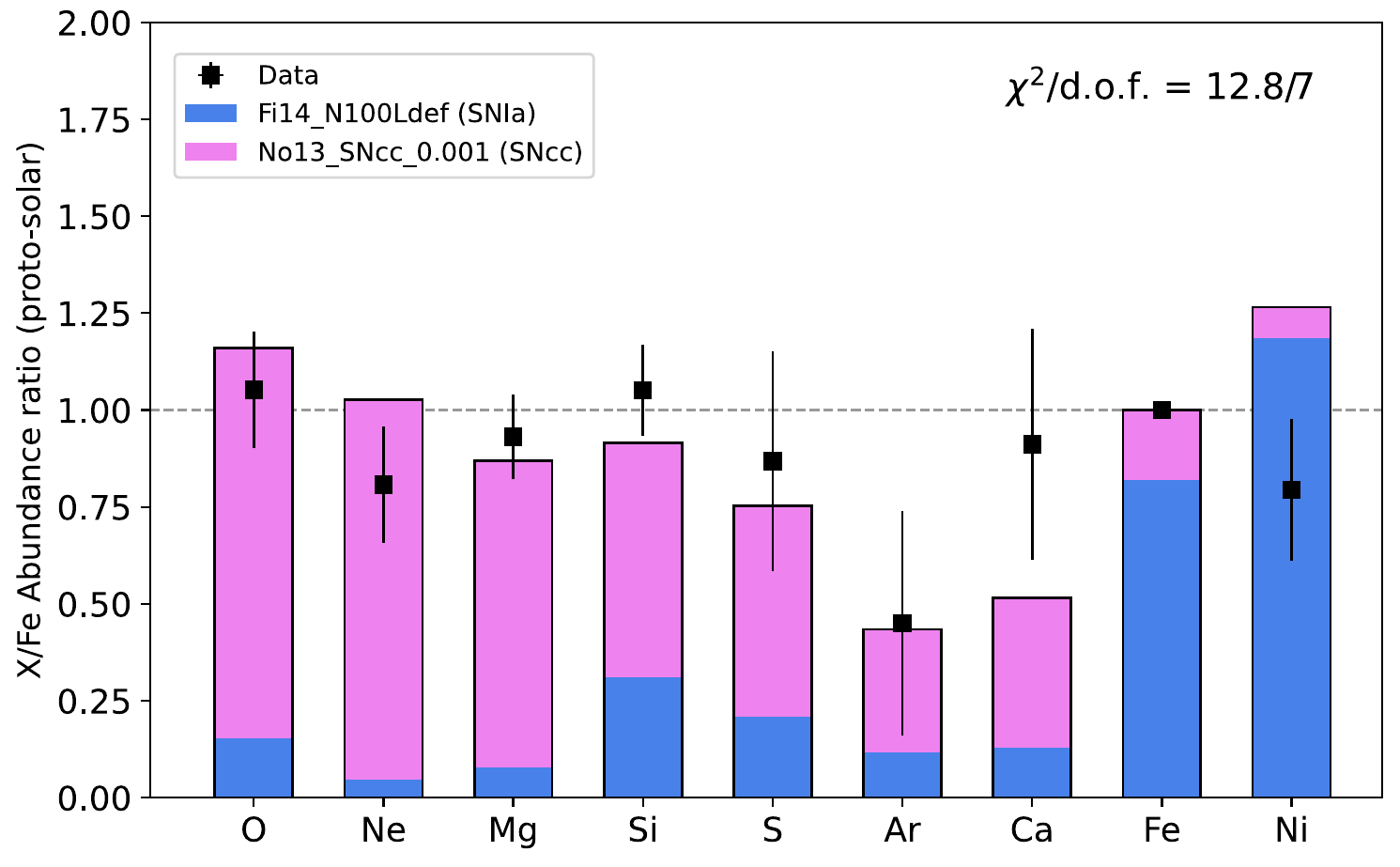}  &  \includegraphics[width=0.5\textwidth]{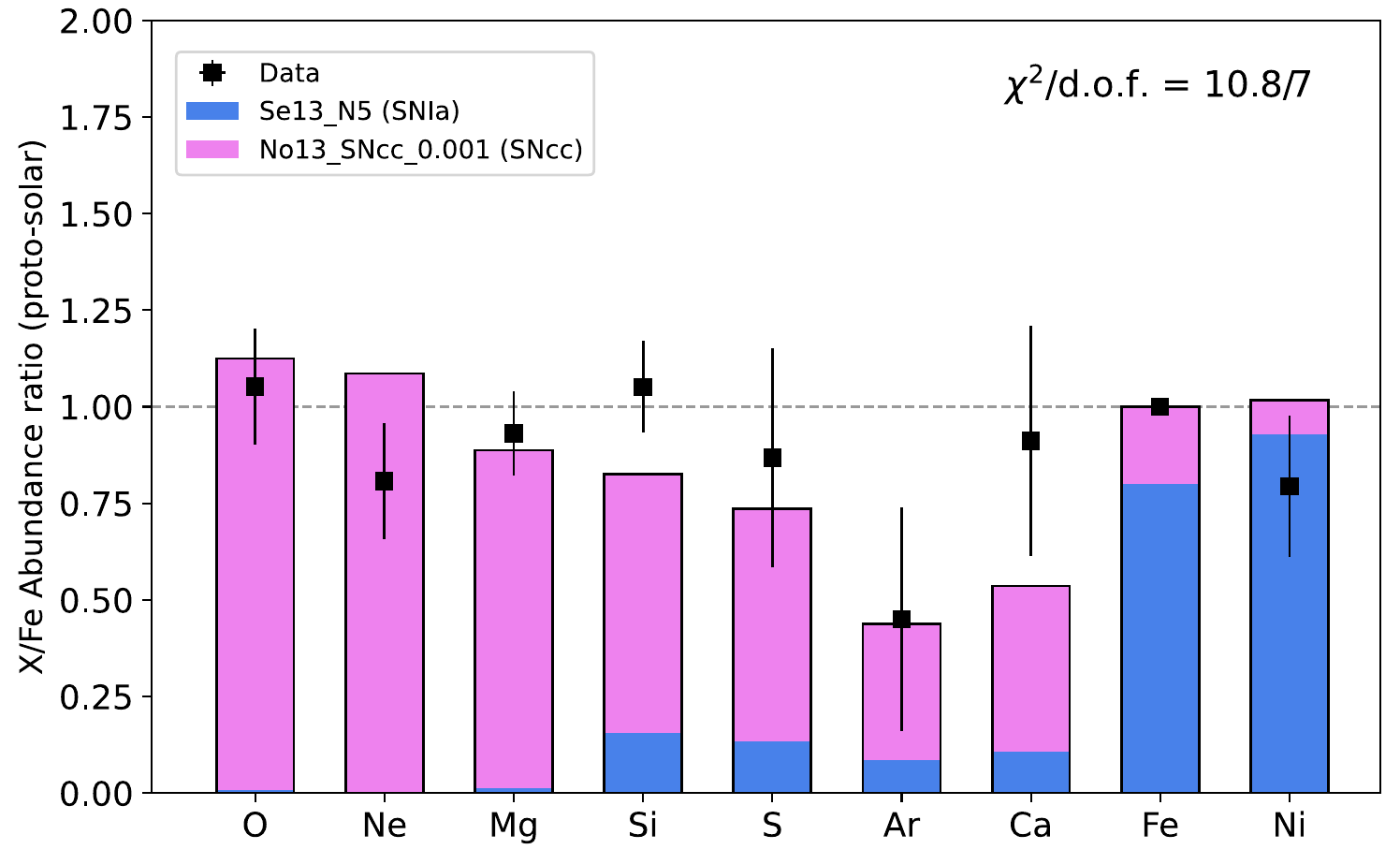}\\
    \end{tabular}
    \begin{tabular}{cc}
       \includegraphics[width=0.5\textwidth]{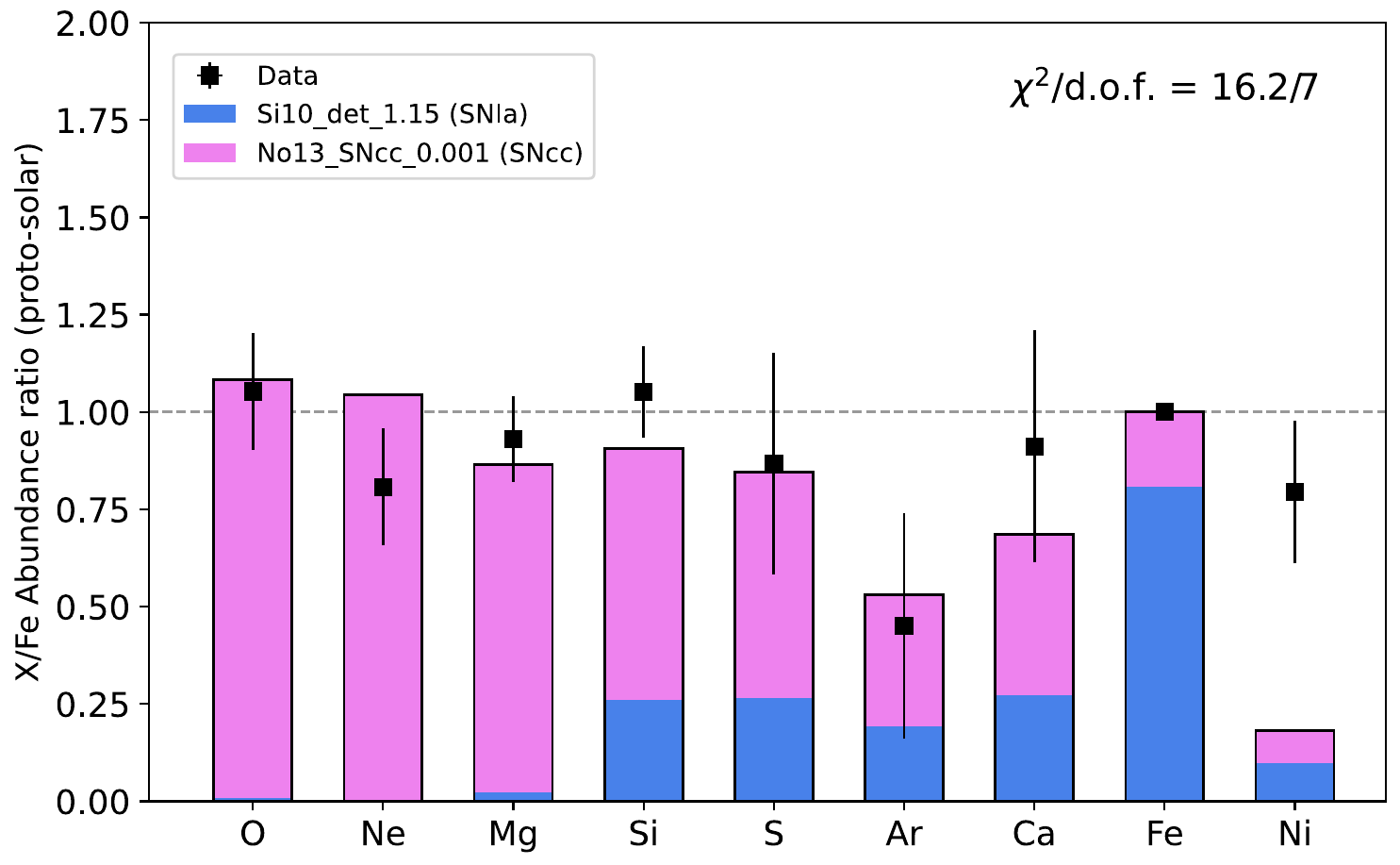}  &  \includegraphics[width=0.5\textwidth]{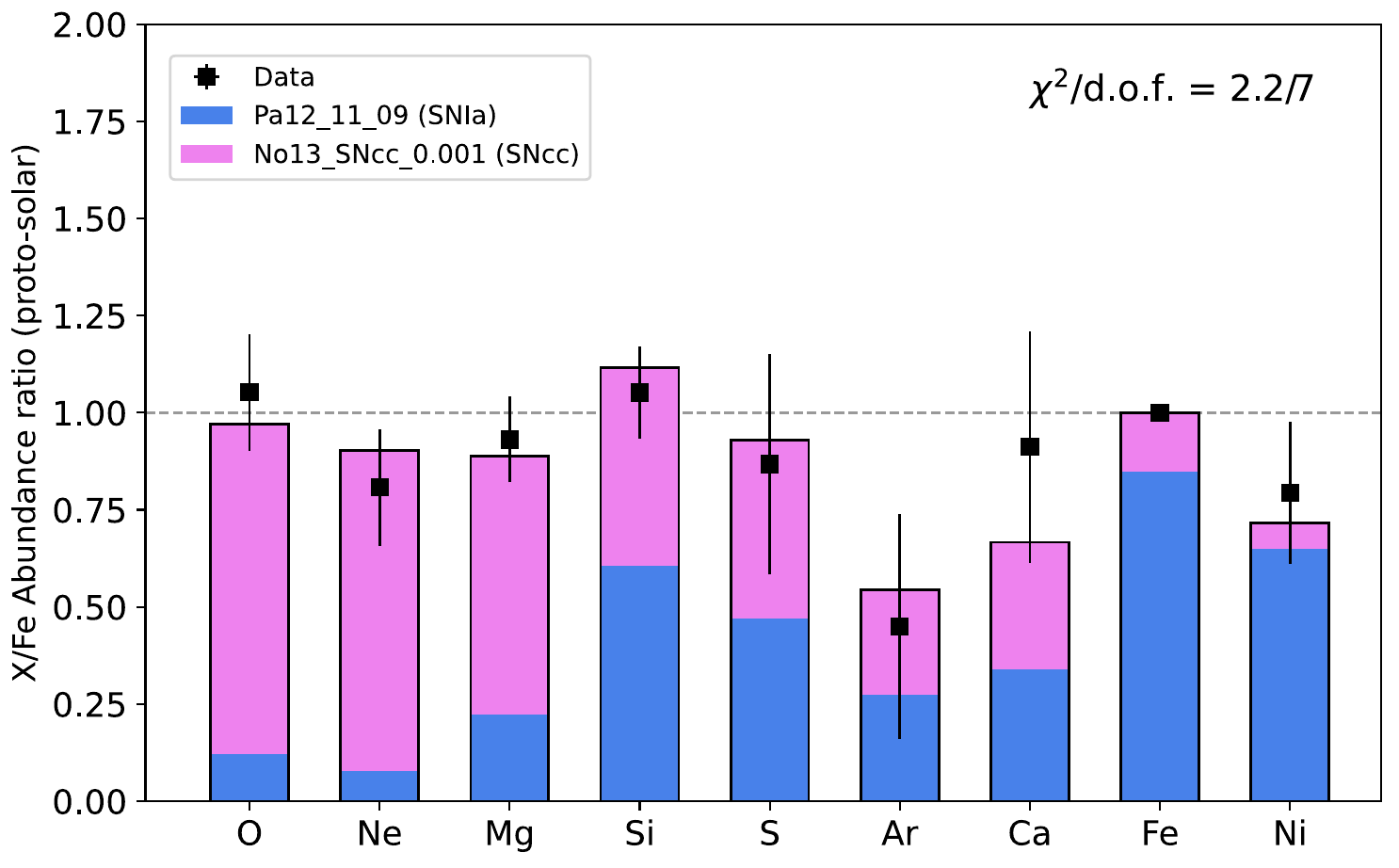}\\
    \end{tabular}
    \caption{Average elemental
    abundance ratios of different elements
    at A2029 center (black data) 
    fitted with SNcc \citep[magenta;][]{2013ARA&A..51..457N} and SNIa (blue)
    yield models. Top-Left: SNIa yield 
    assuming a pure
    deflagration explosion mechanism
    \citep{2014MNRAS.438.1762F},
    top-right: SNIa yield 
    assuming a delayed-detonation
    explosion mechanism
    \citep{2013MNRAS.429.1156S},
    bottom-left: SNIa yield 
    assuming a single-degenerate
    explosion mechanism
    \citep{2010ApJ...714L..52S},
    and bottom-right: SNIa yield 
    assuming a double-degenerate
    explosion mechanism
    \citep{2012ApJ...747L..10P}.
    }
    \label{fig:snia_sncc}
\end{figure*}

We compare the measured X/Fe ratios 
in the core of A2029 with 
predictions from various SNe yield models to draw a clear
picture of the enrichment history of
ICM.
Lighter elements such as O, Ne, and Mg, 
are primarily synthesized in SNcc, 
while heavier elements including 
Ar, Ca, Fe, and Ni are largely
produced by SNIa. 
Thus, the observed X/Fe ratios 
provide a direct and powerful 
diagnostic for disentangling the relative contributions 
of SNcc and SNIa.
For this analysis, 
we constructed composite X/Fe 
ratios at A2029 center—O–Si/Fe derived from RGS, Xtend, Chandra, and EPIC, 
and S-Ni/Fe from Resolve—and fitted
them with a range of SNcc yields 
and four sets of SNIa models within
{\tt abunfit}\footnote{\url{https://github.com/mernier/abunfit}}\citep{2016A&A...595A.126M}.
{ For the SNcc models, we explored several initial metallicity options, including 
0--0.02 solar \citep{2006NuPhA.777..424N,2004ApJ...608..405C} and up to 0.05 solar \citep{2013ARA&A..51..457N}.}
In each case, a Salpeter initial mass function 
(IMF; \citealt{1955ApJ...121..161S})
was assumed.
The four sets of SNIa models considered represent different explosion mechanisms: pure deflagration, 
delayed detonation,
single-degenerate, and
double-degenerate scenarios. 
The resulting best-fit SNe yield combinations, 
along with the observed X/Fe ratios
in the core of A2029, 
are shown in Figure \ref{fig:snia_sncc}.

We find that the SNcc yield model
of \citet{2013ARA&A..51..457N} with 
an initial progenitor metallicity
of $Z_{\rm init}=0.001$ provides
the best match to the 
measured enrichment pattern
in all four fitting cases.
Additionally, for the SNIa contribution, 
the double-degenerate yield 
from detonative 
sub-Chandrasekhar-mass white dwarf 
(WD) explosions \citep{2012ApJ...747L..10P}
reproduces the overall
observed abundance pattern best
in 
the core of A2029,
with $\chi^2/{\rm dof}=2.2/7$ 
(see Figure \ref{fig:snia_sncc})
{ The inferred SNIa-to-total-SNe (SNIa/SNe)
fraction is $\sim$ 30\%.}
In this particular SNIa scenario, 
two WDs with masses of
$1.1\ M_{\odot}$ and $0.9\ M_{\odot}$ 
undergo a violent merger 
(without forming an accretion disk), 
leading to a pure detonation explosion.

We also tested several alternative
SNIa yield models based on
different explosion mechanisms. 
The best-fit 3D pure deflagration SNIa model from 
\citet{2014MNRAS.438.1762F}, when combined with the SNcc model, 
fails to reproduce the observed
Si/Fe and Ca/Fe ratios, yielding a
poor fit with $\chi^2/{\rm dof} = 13.9/7$. 
This model also predicts a 
super-solar Ni/Fe ratio, which is clearly inconsistent with
our measurements in the core of A2029. { Furthermore, it implies a
SNIa/SNe fraction 
of $\sim$45\%}.

Similarly, the best-fit 
3D delayed-detonation SNIa model
by
\citep{2013MNRAS.429.1156S}
better reproduces the observed Ni/Fe
ratio relative to the
deflagration model, but still fails
to reproduce the Si/Fe and Ca/Fe
ratios in the A2029 core, with a 
fit statistic of 
$\chi^2/{\rm dof} = 11.7/7$.
{ The derived SNIa/SNe
fraction is $\sim$ 22\%.}
These findings reinforce earlier 
results by \citet{2016A&A...595A.126M}, 
who showed—using the CHEERS sample 
of 44 clusters and groups—that pure 
deflagration and delayed-detonation SNIa models fail to
reproduce the observed ICM abundance patterns.
However, our analysis with XRISM's significantly improved spectral resolution compared to previous CCD-based instruments,
provides an independent and more precise confirmation of
these conclusions.

We further tested an additional SNIa explosion scenarios: 
the single-degenerate channels.
The best-fit single-degenerate 
model from \citet{2010ApJ...714L..52S},
in which a carbon–oxygen white
dwarf accretes material from a
non-degenerate stellar companion,
fails to reproduce the observed 
Ni/Fe ratio in the core of A2029, 
yielding a poor fit with $\chi^2/{\rm dof} = 17.4/7$. 
{ The best-fit
SNIa/SNe fraction 
is $\sim$27\%.}
This result provides strong 
evidence that the 
single-degenerate channel is
unlikely to be the dominant
progenitor pathway for SNIa 
enrichment in this system.

Among all SNIa yield models considered, the double-degenerate scenario,
when combined with the SNcc yields 
from \citet{2013ARA&A..51..457N} 
better reproduces the observed
abundance pattern at A2029 core.
This result contrasts with the
findings of \citet{2016A&A...595A.126M}, 
who concluded that detonative
sub-Chandrasekhar white dwarf explosions are unlikely to be 
the dominant SNIa channel in
galaxy clusters. 
A key factor contributing to this discrepancy may be the 
super-solar Ni/Fe ratios reported 
in the CHEERS sample, 
which favored deflagration or
hybrid deflagration + delayed-detonation models
that produce larger amounts of Ni.
Measuring Ni abundances
with CCD instruments is 
inherently difficult due to
limited spectral resolution and 
high instrumental background at the energies of Ni–K transitions. 
As a result, those earlier Ni/Fe measurements may carry substantial
systematic uncertainties. 
Our results, based on 
XRISM's high-resolution spectroscopy, 
suggest that the double-degenerate channel as a more 
plausible dominant SNIa 
progenitor channel in A2029.

\subsection{Excess Calcium abundance}

Despite overall good agreement,
as shown in Figure \ref{fig:snia_sncc}, the double-degenerate SNIa yield 
model from \citet{2012ApJ...747L..10P}, 
combined with the SNcc model from \citet{2013ARA&A..51..457N},
struggles to reproduce the
observed Ca/Fe ratio in the 
core of A2029.
Similar, elevated Ca/Fe ratios
relative to standard SN yield predictions have 
been reported in several previous studies
\citep[e.g.,][]{2006A&A...449..475W,2007A&A...465..345D,2015A&A...575A..37M,2016A&A...592A.157M}.
One possible explanation for the 
excess Ca in the ICM is the contribution from
Ca-rich gap transients, as proposed by
\citet{2014ApJ...780L..34M}.
These events are identified as peculiar 
Type Ib/Ic SNe
\citep[e.g.,][]{2017ApJ...836...60L}
and are believed to originate from helium-accreting white dwarfs 
rather than core-collapse progenitors
\citep[e.g.,][]{2011ApJ...738...21W,2015MNRAS.452.2463F,2023ApJ...944...22Z}.
Ca-rich gap transients are characterized by
calcium-dominated nebular spectra
and unusually high photospheric expansion velocities 
\citep{2012ApJ...755..161K}.
These Ca-rich gap transients tend to
explode far from their
host galaxies, 
making it easier for their ejecta
to escape and become well mixed 
into the surrounding ICM
\citep[e.g.,][]{2013MNRAS.432.1680Y}.

\begin{figure}
    \centering
\includegraphics[width=0.5\textwidth]{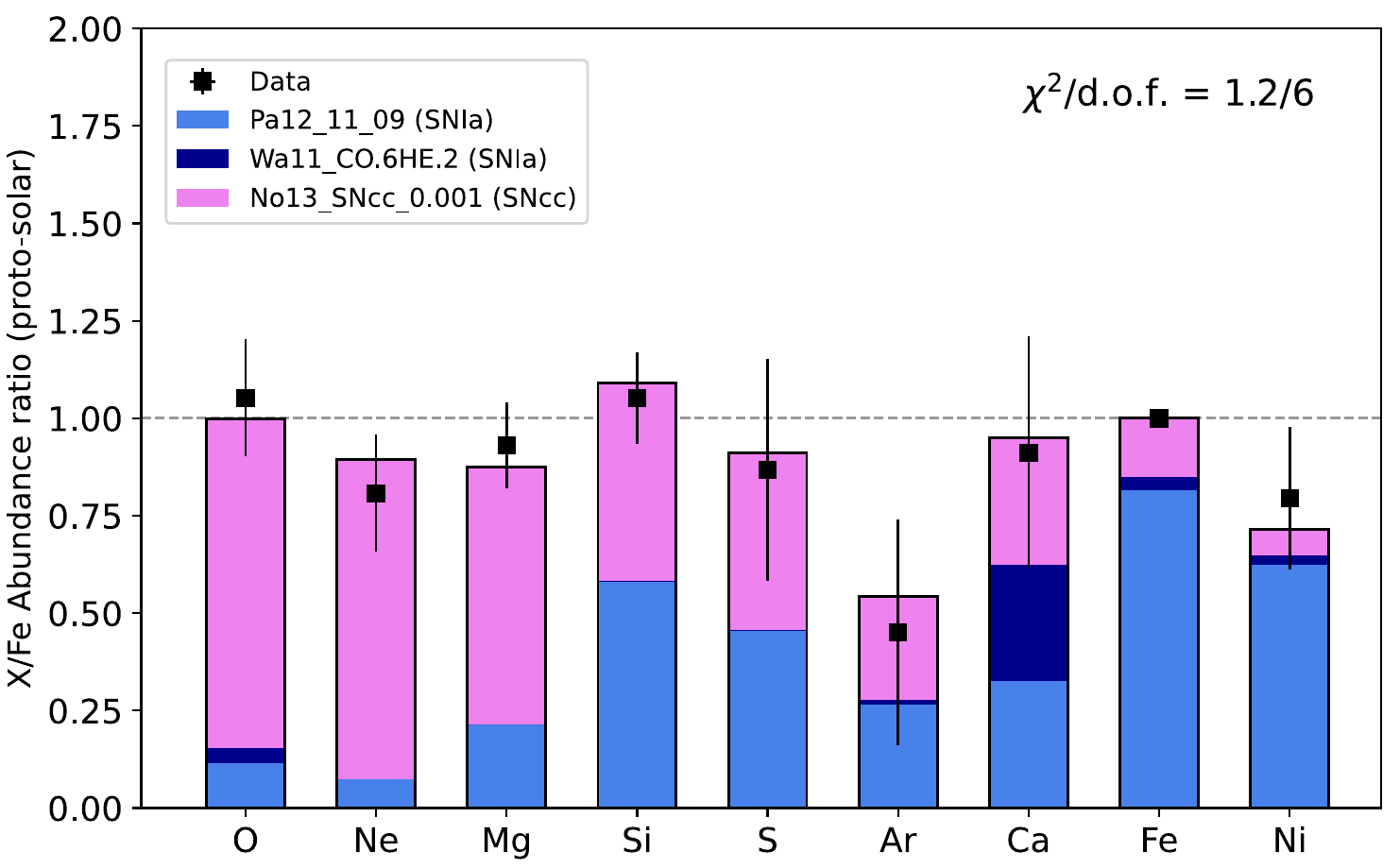} 
    \caption{Average elemental
    abundance ratios of different elements
    at A2029 center (black data) 
    fitted with SNcc \citep[magenta;][]{2013ARA&A..51..457N} and SNIa yield 
    assuming a double-degenerate
    explosion mechanism
    \citep{2012ApJ...747L..10P} and Ca-gap transients \citep{2011ApJ...738...21W}.
    }
    \label{fig:ca_gap}
\end{figure}

Motivated by this,
we tested a model that adds the 
Ca-rich gap transient yield from \citet{2011ApJ...738...21W} to 
our best-fit combination of the \citet{2012ApJ...747L..10P} SNIa
and \citet{2013ARA&A..51..457N}
SNcc models.
As shown in Figure \ref{fig:ca_gap}, 
adding a Ca-rich gap supernova component improves the match to the observed Ca/Fe ratio.
This modification results 
in a noticeably better overall fit, with an F-test p-value
of approximately 0.06 
when compared to the one SNIa (double-degenerate) + one SNcc model. 
Our results clearly indicate a 
complex chemical enrichment history 
in the core of A2029.
Hitomi observations of the Perseus
Cluster \citep{2019MNRAS.483.1701S} 
also revealed elevated Ca abundances, which have been
interpreted as possible evidence
for contributions from
Ca-rich gap transients.

Finally, we note that SNe yield models
suffer from significant uncertainties
up to a factor of 2 \citep{2009MNRAS.399..574W}, therefore
the excess Ca/Fe at A2029 center 
should be interpreted with caution.
We need more deep XRISM observations 
of galaxy clusters and higher quality SNe yield models to
further explore ICM enrichment.
{ Additionally, the derived SNe yield models represent effective, time-averaged mixtures rather than exact combinations of individual SNcc and SNIa events. In reality, the enrichment history involves evolving IMFs, progenitor masses, and explosion channels over cosmic time, making the true picture more complex than the fitted model implies.}

\section{Summary}\label{sec:summary}
We have measured chemical abundance pattern
of several
$\alpha$ and Fe-peak elements within 
the core of A2029 ($3'\times3'$) using XRISM and archival
observations from XMM-Newton and Chandra.
Our main findings are summarized below.

\begin{itemize}
    \item[\bf --] The two-temperature (2T CIE) model fit from Resolve shows cooler and hotter gas components that are consistent with those measured by Xtend and EPIC. Although RGS is limited to the 0.2–2 keV band and is more sensitive to hotter gas, its measured 
    hotter gas temperature (4.40$^{+0.11}_{0.10}$ keV) 
    is consistent with the cooler components found by the other instruments.

    \item[\bf --] 
    We have constructed an average X/Fe pattern 
    for the A2029 core by combining
    measurements from multiple instruments.
    For the S/Fe, Ar/Fe, Ca/Fe, and Ni/Fe 
    abundance ratios, we primarily leverage Resolve’s superior spectral
    resolution and sensitivity.
    Abundance ratios for O/Fe and Ne/Fe are
    averaged from
    Xtend+RGS and Xtend+RGS+EPIC+Chandra, respectively. 
    Mg/Fe and Si/Fe ratios are averaged from EPIC+Chandra and Xtend+EPIC+Chandra, respectively.

    \item[\bf --] By comparing the average
    X/Fe ratios with
    various SNcc and SNIa
    yield models, we find that SNcc yield model
    from \citet{2013ARA&A..51..457N}, assuming
an initial progenitor metallicity
of $Z_{\rm init}=0.001$, combined with a SNIa model involving sub-$M_{\rm Ch}$ WD detonation 
in a double-degenerate system \citep{2012ApJ...747L..10P}, provides the best overall match to the observed abundance pattern in the cluster core.

    \item[\bf --] We observe an excess in Ca abundance in the core of A2029 that cannot be reproduced by the double-degenerate SNIa yield model.  
When we include a Ca-rich gap component
  \citep{2011ApJ...738...21W} in the SNe modeling, the agreement with the observed Ca/Fe ratio significantly improves.
    
\end{itemize}

\section*{Acknowledgement}
We sincerely thank the anonymous referee for their helpful suggestions and comments.
We gratefully acknowledge the dedicated efforts of the many engineers and scientists whose hard work over the years made the XRISM mission possible.
AS and EDM acknowledges support from NASA grants 80NSSC20K0737 and 80NSSC24K0678.
MS acknowledges support from NASA grant 80NSSC23K0650.   LL acknowledges the financial contribution from the INAF grant 1.05.12.04.01. S.E. acknowledges the financial contribution from {\it Theory Grant / Bando INAF per la Ricerca Fondamentale 2024}  on ``Constraining the non-thermal pressure in galaxy clusters with high-resolution X-ray spectroscopy'' (1.05.24.05.10).
This paper employs a list of Chandra datasets, obtained by the Chandra X-ray Observatory, contained in~\dataset[DOI: 10.25574/cdc.499]{https://doi.org/10.25574/cdc.499}

\bibliographystyle{aasjournal}
\bibliography{sample631} 

\end{document}